\documentclass[%
groupedaddress,
 amsmath,amssymb,
 aps,
prb,
twocolumn,
floatfix,
notitlepage
]{revtex4-1}

\usepackage{graphicx}% Include figure files
\usepackage{dcolumn}% Align table columns on decimal point
\usepackage{bm}% bold math
\usepackage{subfigure}%subfigure enviroment
\makeatletter
\renewcommand{\thesubfigure}{\alph{subfigure}}% (a) -> a
\renewcommand{\@thesubfigure}{\thesubfigure)\hskip\subfiglabelskip}% a -> a)
\makeatother
\usepackage[colorlinks=true,bookmarks=true,%
     pdfborder={0 0 0},linkcolor=blue,urlcolor=blue,%
     citecolor=blue,
     breaklinks=true,bookmarksnumbered=true]{hyperref}
\def \k{\bm{k}}

\def \pt{\partial}

\def \Ree{\mathcal{R}e}
\def \Imm{\mathcal{I}m}

\def \beq{\begin{equation}}
\def \eeq{\end{equation}}
\def \beqarr{\begin{eqnarray}}
\def \eeqarr{\end{eqnarray}}
\def \bspt{\begin{split}}
\def \espt{\end{split}}
\def \bef{\begin{figure}}
\def \enf{\end{figure}}

\def \bpm{\begin{pmatrix}}
\def \epm{\end{pmatrix}}
\def \sgn{\text{Sgn}}
\newcommand{\abs}[1]{\lvert#1\rvert}

\newcommand{\refdisp}[1]{Ref.~[\onlinecite{#1}]}

\begin{document}

%\preprint{APS/123-QED}

\title{A Strange Metal from Gutzwiller correlations in infinite
dimensions II: Transverse Transport, Optical Response and Rise of Two Relaxation Rates
}% Force line breaks with \\

 \author{ Wenxin Ding$^1$, Rok \v{Z}itko $^{2,3}$, and B Sriram Shastry$^1$}

\affiliation{
$^1$Physics  Department, University of California, Santa Cruz, California, 95060,\\ $^2$
Jo\v{z}ef Stefan Institute, Jamova 39, SI-1000 Ljubljana, Slovenia\\$^3$Faculty for Mathematics and Physics, University of
Ljubljana, Jadranska 19, SI-1000 Ljubljana, Slovenia
 }%

\date{\today}% It is always \today, today,

\begin{abstract}
Using two approaches to strongly correlated systems, the extremely
correlated Fermi liquid theory  and the dynamical mean field
theory, we compute the transverse transport coefficients,
namely the Hall constants $R_H$ and Hall angles $\theta_H$, and
the longitudinal and transverse optical response of the $U=\infty$
Hubbard model in the limit of infinite dimensions. {We focus} on  two {successive}
low-temperature regimes, the Gutzwiller correlated Fermi liquid (GCFL)
and the Gutzwiller correlated strange metal (GCSM). We find that the
Hall angle $\cot \theta_H$ is proportional to $T^2$ in the GCFL regime, 
while on warming into the GCSM regime it first passes through a downward bend 
and then continues as $T^2$.
Equivalently, $R_H$ is
weakly temperature dependent in the GCFL regime, but becomes strongly
temperature dependent in the GCSM regime. Drude peaks are found for both the
longitudinal optical conductivity $\sigma_{xx}(\omega)$ and the
optical Hall angles $\tan \theta_H(\omega)$ below certain
characteristic energy scales. By comparing the relaxation rates
extracted from fitting to the Drude formula, we find that in the GCFL
regime there is a single relaxation rate controlling both longitudinal
and transverse transport, while in the GCSM regime two different
relaxation rates emerge. We trace the origin of this behavior
to the dynamical particle-hole asymmetry of the Dyson self-energy,
{arguably} a generic feature of doped Mott insulators.
\end{abstract}

%\begin{description}
% \end{description}

\pacs{Valid PACS appear here}% PACS, the Physics and Astronomy
\maketitle
\section{Introduction}

In a recent study \cite{Ding2017} we have presented results for the
longitudinal resistivity and low-temperature thermodynamics of the
Hubbard model (with the repulsion parameter $U = \infty$) in the infinite dimensional limit. In this limit, we can obtain the complete single-particle Green's functions using two methods: the dynamic mean field theory (DMFT) \cite{Vollhardt1989,Georges1996,Deng2013,Xu2013d%Badmetal,HFL
}, and the extremely correlated Fermi liquid (ECFL) theory\cite{Perepelitsky2016, Shastry2011}, with some overlapping results and comparisons in Ref. [\onlinecite{Rok}].
 These studies capture the non-perturbative local Gutzwiller
 correlation effects on the longitudinal resistivity {$\rho_{xx}$}
 quantitatively\cite{Deng2013,Xu2013d,Perepelitsky2016}. A recent
 study by our group addresses the physically relevant case of two dimensions\cite{Shastry-Mai}, with important results for many variables discussed here.

The present work extends the study of Ref.
[\onlinecite{Ding2017}]{, using the ECFL scheme of Ref.
[\onlinecite{Perepelitsky2016}]}, to the case of the Hall conductivity
{$\sigma_{xy}$} and the finite frequency (i.e. optical)
conductivities. One goal is to further test ECFL with the exact DMFT
results for these quantities which are more challenging to calculate.
More importantly, however, by combining the various calculated
conductivities we are able to uncover the emergence of two different
transport relaxation times. In cuprate superconductors, various
authors \cite{Chien1991, Anderson1991,Ando,Boebinger,NCCO} have
commented on the different temperature ($T$) dependence of the
transport properties in the normal phase. The cotangent Hall
angles{, defined as the ratio of the longitudinal conductivity
$\sigma_{xx}$ and the Hall conductivity, $\cot(\theta_H) =
\sigma_{xx}/\sigma_{xy}$,} is close to quadratic as in conventional
metals. Meanwhile, the longitudinal resistivity has unusual linear
temperature dependence \cite{Ando2004b}. Understanding the ubiquitous
$T^2$ behavior of $\cot(\theta_H)$ in spite of the unconventional
temperature dependence of the longitudinal resistivity is therefore
quite important.

In Ref. [\onlinecite{Ding2017}] we found that at the lowest
temperatures the system is a {\em Gutzwiller-correlated Fermi liquid}
(GCFL) with $\rho_{xx} \propto T^2$. Upon warming one finds a regime
with linear temperature dependence of the resistivity
$\rho_{xx}$\cite{Ding2017}, which is reminiscent of the {\em strange
metal} regime in the cuprate phase diagrams \cite{Ando2004b}. It is
termed the {\em Gutzwiller-correlated strange metal} (GCSM)
regime\cite{Ding2017}. Previous studies\cite{Deng2013,Xu2013d}
established the GCFL and GCSM regimes using the longitudinal
resistivity. Here we focus instead on the Hall constants {$R_H =
\sigma_{xy}/\sigma_{xx}^2$} and the Hall angles\cite{Xu2013d}, as well
as on the optical conductivity\cite{Deng2013} and optical Hall angles.
In the GCFL regime, the primary excitations are coherent
quasiparticles that survive the Gutzwiller correlation, and there is a
single transport relaxation time, as one would expect for a
conventional Fermi liquid. Upon warming up into the GCSM regime, the
longitudinal and transverse optical scattering rates become different.
It appears that the existence of two separate scattering times is a
generic characteristic of the GCSM regime.

This work is organized as follows. First we {summarize the Kubo
formulas} used to calculate the transport coefficients in Sec.
(\ref{sec:formalism}). We then revisit in Sec.
(\ref{subsec:boltzmann}) the familiar Boltzmann transport theory from
which two separate relaxation times can be naturally derived. The
results for the $dc$ transport properties are presented in Sec.
(\ref{sec:dc}) and those of optical conductivities in Sec.
(\ref{sec:optical}). In Sec. (\ref{sec:analysis}) we interpret the two
scattering times found in the GCSM regime through the particle-hole
asymmetry of dynamical properties (spectral function) of the system.
In conclusion we discuss the implication of this work for strongly
correlated matter.

\section{Kubo formulas}\label{sec:formalism}

The transport properties of correlated materials can be easily evaluated in the limit of infinite dimensions because the vertex corrections are absent\cite{Khurana1990}.
For dimensions $d>3$, the longitudinal conductivity $\sigma_{xx}$ is straightforwardly generalized as the electric field remains a $d$-dimensional vector.
The generalization is less clear for the transverse conductivity and Hall constants, because the magnetic field is no longer a vector but rather a rank-2 tensor defined through the electromagnetic tensor. Nevertheless,  $\sigma_{xy}$ can still be defined through suitable current-current correlation functions.

The input to the transport calculation is the single-particle Green's function $G(\omega, \k)$, calculated in the following within either ECFL or DMFT.
The Kubo formulas can be written as \cite{Jarrell1993, Voruganti1992}
\begin{eqnarray}\label{eq:sxx}
  & \sigma_{xx} = 2 \pi q_e^2\sum_k \Phi_k^{xx} \int d\omega (-\frac{\pt f(\omega)}{\pt \omega}) \rho^2_G(\omega,\k),\\
\label{eq:sxy}
  & \sigma_{xy}/B = \frac{4 \pi^2 q_e^3}{3} \sum_k \Phi_k^{xy} \int d\omega (-\frac{\pt f(\omega)}{\pt \omega}) \rho^3_G(\omega,\k),
\end{eqnarray}
where $\rho_G(\omega,\k) = -\Imm G(\omega,\k) / \pi$ is the single-particle spectral function and $q_e= - |e|$ is the electron charge. $\Phi_k^{xx} = (\epsilon_k^x)^2$ and $\Phi_k^{xy} = (\epsilon_k^y)^2 \epsilon_k^{xx} - \epsilon_k^y \epsilon_k^x \epsilon_k^{xy}$ are called transport functions, with
$\epsilon^\alpha_k = \pt \epsilon_k/\pt k_{\alpha}$ and
$\epsilon^{\alpha\beta}_k = \pt^2 \epsilon_k/\pt k_{\alpha} \pt
k_{\beta}$, $\epsilon_k$ being the energy dispersion.
We set $\hbar$ to 1.

%{\it Approximation for $k$-Integration:}

It is more convenient to convert the multi-dimensional $k$-sums into
energy integrals:
\begin{eqnarray}\label{eq:sxx-ep}
  & \sigma_{xx} = \sigma_{0} 2 \pi D \int d\epsilon \frac{\Phi^{xx}(\epsilon)}{\Phi^{xx}(0)} \int d\omega (-\frac{\pt f(\omega)}{\pt \omega}) \rho^2_G(\omega, \epsilon),\\
\label{eq:sxy-ep}
  & \sigma_{xy}/B = \sigma_{0} \frac{4 \pi^2 D q_e}{3} \int d\epsilon \frac{\Phi^{xy}(\epsilon)}{\Phi^{xx}(0)} \int d\omega (-\frac{\pt f(\omega)}{\pt \omega}) \rho^3_G(\omega, \epsilon),
\end{eqnarray}
where $\Phi^{xx(xy)}(\epsilon) = \sum_{\k} \Phi_{\k}^{xx(xy)} \delta(\epsilon - \epsilon_{\k})$,  $\sigma_{0} = q_e^2 \Phi^{xx}(0)/D$ is the Ioffe-Regel-Mott conductivity, $D$ is half-bandwidth, and $\rho_G(\omega, \epsilon) = \rho_{G}(\omega, \k)$ such that $\epsilon=\epsilon_k$.
In $d$ dimensions the transport functions on the Bethe lattice
are \cite{Arsenault2013}
\begin{eqnarray}
  \Phi^{xx}(\epsilon) = \frac{1}{3d} (D^2 - \epsilon^2) \rho_0(\epsilon) , \label{eq:tr-fun-xx}\\
  \Phi^{xy}(\epsilon) = - \frac{1}{ 3d(d-1)} \epsilon (D^2 - \epsilon^2) \rho_0(\epsilon) ,\label{eq:tr-fun-xy}
\end{eqnarray}
where $\rho_0(\epsilon) = \frac{2}{\pi D^2} \sqrt{D^2 - \epsilon^2} \Theta(D-|\epsilon|)$ is the non-interacting density of states on the Bethe lattice and $D$ is the half bandwidth. Even though the transport function results indicate that $\sigma$ vanishes as $d\rightarrow \infty$, we can redefine the conductivities in this limit as the sum of all components: $\sigma_{L} = \sum_\alpha \sigma_{\alpha \alpha}$, $\sigma_{T} = \sum_{\alpha \neq \beta} \sgn[\alpha - \beta] \sigma_{\alpha \beta}$ with $\alpha (\beta) = 1,2,\dots, d$.
More importantly, the $d$-dependence directly drops out when we compute the Hall constant $R_H =\sigma_{xy} / \sigma_{xx}^2$. For the rest of this work, we shall re-define  $\sigma_{xx}$ and $\sigma_{xy}$ via $\sigma_{L}$ and $\sigma_{T}$ considering that all components of $\sigma_{L(T)}$ are equal so that both the $d$-factor and the constant factor drop out from the transport functions:
\begin{eqnarray}
  \label{eq:re-define-sigma}
  \sigma_{xx} = 3 \sigma_{L}, \qquad \Phi^{xx}(\epsilon) =  (D^2 - \epsilon^2) \rho_0(\epsilon),\\
  \sigma_{xy} = 3 \sigma_{T}, \qquad \Phi^{xy}(\epsilon) = - \epsilon (D^2 - \epsilon^2) \rho_0(\epsilon).
\end{eqnarray}

\section{Two-relaxation-time behavior in the Boltzmann theory}\label{subsec:boltzmann}

In Boltzmann theory, the transport properties can be obtained by solving for the distribution function in the presence of external fields from the Boltzmann equation\cite{Ziman1960}:
\begin{equation}\label{eq:l-Boltzmann}
  \frac{\pt \, \delta \! f}{\pt t} - \frac{q_e}{\hbar c} \bm{v} \times\bm{B} \cdot \frac{\pt \, \delta \! f}{ \pt \k} + \bm{v} \cdot q_e \bm{E}(t)\left(-\frac{\pt f^0}{\pt \epsilon}\right) = \hat{L} \, \delta \! f,
\end{equation}
where $f$ is the full distribution function that needs to be solved, $f^0$ is the Fermi-Dirac distribution function, $\delta \! f=f-f^0$, and $ \hat{L} \delta \! f$ represents the linearized collision integrals.

In the regime of linear response, we expand $\delta \! f^{E,B}$ in powers of the external fields to second order as
\begin{equation}
  \delta \! f^{E,B} = \delta \! f^{E,0} + B \delta \! f^{E,1},
\end{equation}
where $\delta \! f^{E,0}$ is the solution in the absence of magnetic fields, and both $\delta \! f^{E,0}$ and $\delta \! f^{E,1}$ are linear in $E$.
In the relaxation-time-approximation (RTA)\cite{Feng2005} we replace the collision integrals as $\hat{L}_{\k} \delta \! f \rightarrow  - \delta \! f/\tau$ where $\tau$ is assumed to be $\k$-independent.
However, $\hat{L} \delta \! f^{E,0}$ and $\hat{L} \delta \! f^{E,1}$
are in principle governed by different relaxation times, as pointed out by Anderson \cite{Chien1991, Anderson1991}.
Writing
\begin{equation} \label{eq:tauH-RTA}
\hat{L} \delta \! f^{E,0} \rightarrow - \frac{\delta \! f^{E,0}}{\tau_{tr}}, \qquad
\hat{L} \delta \! f^{E,1} \rightarrow - \frac{\delta \! f^{E,1}}{\tau_{H}},
\end{equation}
we obtain

\begin{align}
&\sigma_{xx}(\omega) = \frac{\omega_p^2}{4 \pi} \frac{\tau_{tr}}{1 - i \omega \tau_{tr}},\\
&\sigma_{xy}(\omega)/B = \frac{\omega_p^2 \omega_c/B }{4 \pi^2}\frac{\tau_{H}}{1 -  i \omega \tau_{H}} \frac{\tau_{tr}}{1 - i \omega \tau_{tr}},
  \label{eq:Boltzmann-sigma-xy}
\end{align}
where
\begin{align}
&\frac{\omega_p^2}{4 \pi} = \int \frac{d^dk}{(2 \pi)^d} 2 q_e^2 v_x^2 (-\pt_\epsilon f^0),\\
 & \frac{\omega_c}{B} = \omega_p^{-2} \int \frac{d^dk}{(2 \pi)^d} 2 q_e^3 (v_x^2\pt_{k_y}v_{y} - v_x v_y \pt_{k_x} v_y) \pt_\epsilon f^0 ,\label{eq:omega-c}
\end{align}
$v_a = \pt_{k_a} \epsilon(\k)$ is the velocity in direction $a$, $\epsilon(\k)$ is energy dispersion of the electrons and ${\bf B} = \hat{z} B$. Then the Hall angle is
\begin{equation}
  \label{eq:theta_H_drude}
  \tan \theta_H(\omega) = \frac{\omega_c}{\pi} \frac{\tau_{H}}{1 -  i \omega \tau_{H}}.
\end{equation}
Therefore, the optical conductivities can be cast in the Boltzmann-RTA form as
\begin{align}
 \frac{\sigma_{xx}(0)}{\Ree [\sigma_{xx}(\omega)]} &= 1 + \omega^2 \tau_{tr}^2, \label{eq:sigma-xx-tau}\\
  \frac{\sigma_{xy}(0)/B}{\Ree [\sigma_{xy}(\omega)]/B} &= 1 + \omega^2 (\tau_{tr}^2 + \tau_H^2) + \tau_{tr}^2 \tau_H^2 \omega^4\label{eq:sigma-xy-tau},\\
  \frac{ \theta_H(0)}{\Ree[\theta_H(\omega)]} & = 1 + \omega^2 \tau_H^2. \label{eq:Ha-tau}
\end{align}

  The $dc$ and $ac$ transport coefficients of a microscopic theory do {\it not} necessarily take the form of the Boltzmann RTA theory. In the rest of this work, we study both the $dc$ and the real part of the $ac$ transport coefficients, and consider them as
  \begin{align}
&\Re e [\sigma_{xx}(\omega)] =  \frac{\sigma_{xx}(0)}{1 + \tau^2_{tr} \omega^2 + \mathcal{O}(\omega^4)},\\
  &  \Re e[\tan \theta_H(\omega)/B] = \frac{\tan \theta_H(0)/B}{1 + \tau_{H} ^2 \omega^2 + \mathcal{O}(\omega^4)}.
  \label{eq:sigma-w}
  \end{align}
  The relaxation times $\tau_{tr}$ and $\tau_H$ are extracted from the
  low frequency part of $\Re e [\sigma_{xx}(\omega)]$ and $\Re e[\tan
  \theta_H(\omega)/B]$ by fitting to the above expressions. Although
  computing $\Ree[\theta_H(\omega)]$ requires both real and imaginary
  parts of the optical conductivities,  we can make the approximation $\Ree[\theta_H(\omega)] \simeq \Ree [\sigma_{xy}(\omega)] / \Ree [\sigma_{xx}(\omega)]$ when $\omega$ of concern is small. We expect $\tau_{tr}$ and $\tau_H$ to have similar temperature and density dependence as $\sigma_{xx}(0)$ and $\tan \theta_H(0)/B$.

\section{$dc$ Transport}\label{sec:dc}

\begin{figure}
  \centering
  %\subfigure[]{\label{subfig:Rxx}
    \includegraphics[width=8cm]{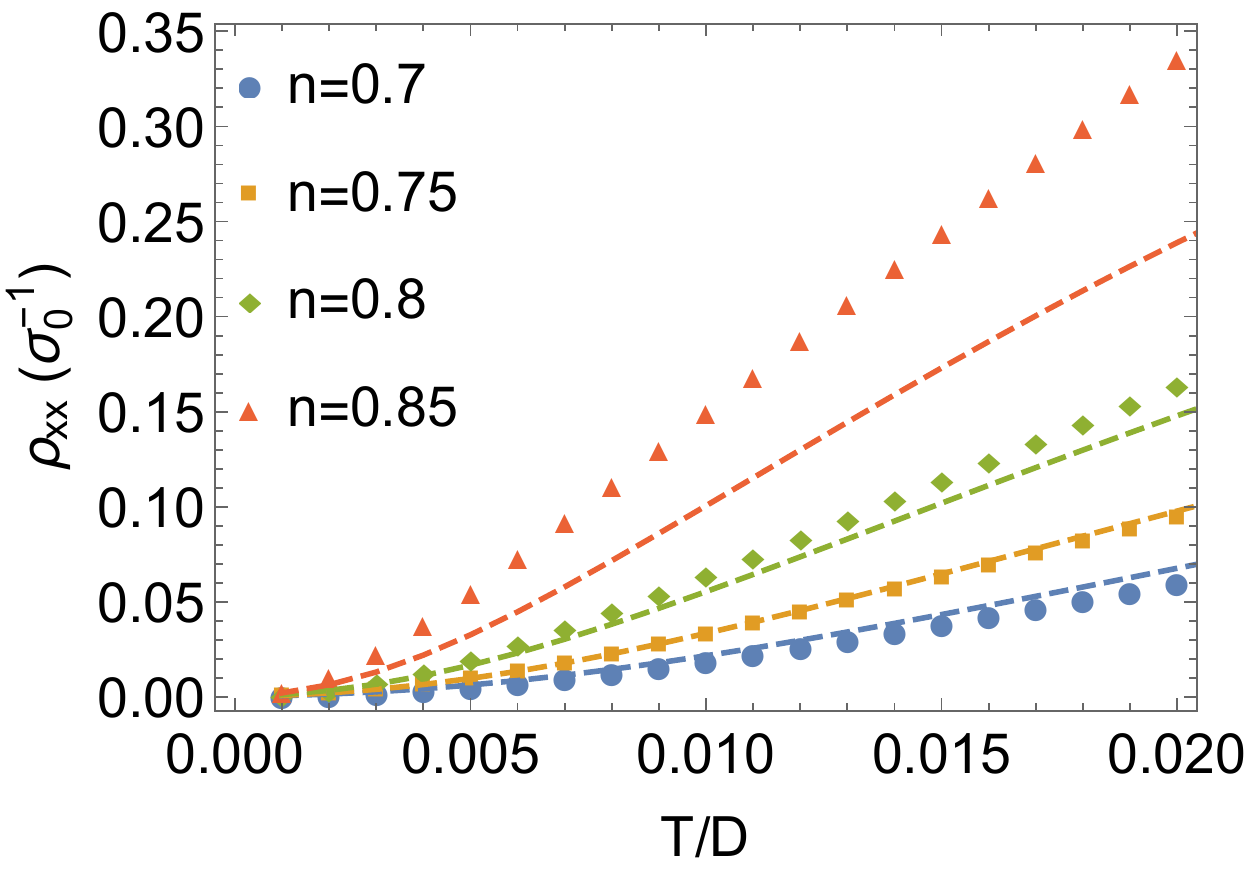}%}
  %\subfigure[]{\label{subfig:Rxx-rescale}\includegraphics[width=8cm]{./rhoxx-rescale.pdf}}
 \caption{Temperature dependence of the $dc$ resistivity $\rho_{xx}$ of the $U=\infty$ Hubbard model from DMFT (dashed lines) and ECFL (solid symbols) for a range of electron densities $n$. The horizontal axis corresponds to absolute temperatures. }
  \label{fig:rhoxx0}
\end{figure}

We now use the Kubo formulas
%ulas Eq. (\ref{eq:sxx}) and (\ref{eq:sxy})
to compute the transport coefficients within the ECFL and DMFT approaches.
We plot the ECFL results as solid symbols and the DMFT results as dashed lines using the same color for each density unless specified otherwise. As we shall demonstrate, the agreement between the DMFT and ECFL results follows the same qualitative trend for all quantities considered: it is better at lower temperatures, lower frequencies, and at lower density (higher hole doping).

\begin{figure*}[t]
  \centering
   \subfigure[]{\label{subfig:RH}\includegraphics[width=.75 \columnwidth]{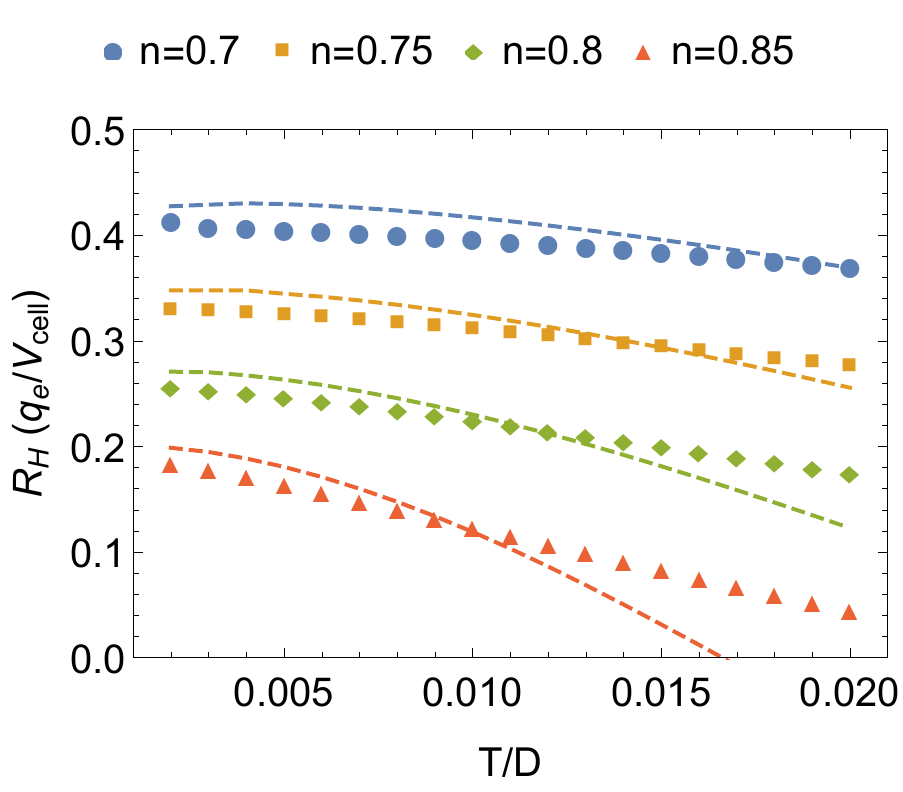}}
   \subfigure[]{\label{subfig:nHA}\includegraphics[width=.75 \columnwidth]{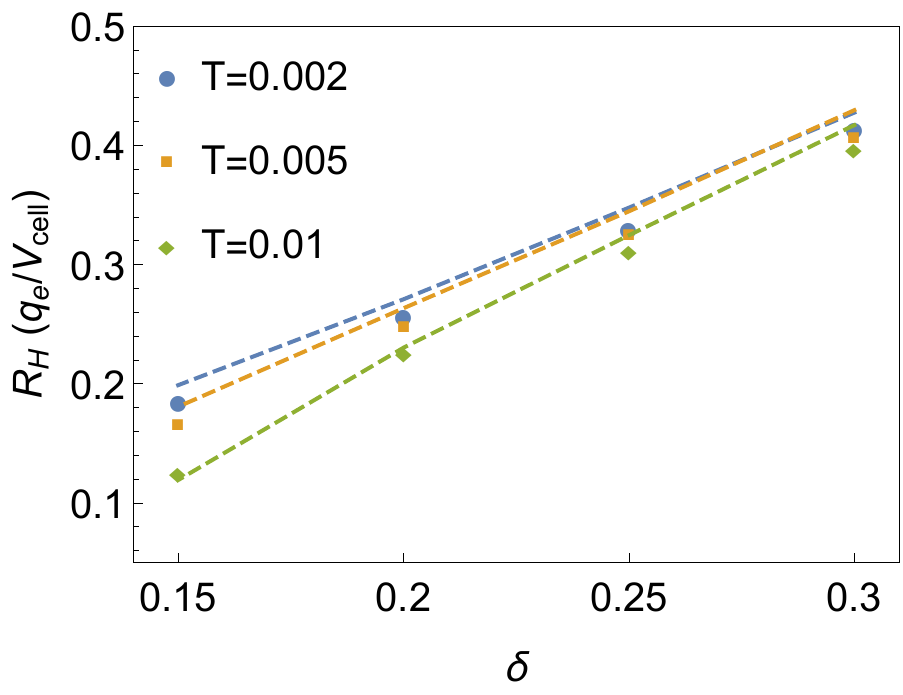}}
   \caption{Temperature dependence of the Hall constants $R_H$ (\ref{subfig:RH}) and $R_H$ at $T= 0.002D, 0.005D, 0.01D$ as functions of the hole density $\delta = 1-n$ (\ref{subfig:nHA}) for both DMFT (dashed lines) and ECFL (solid symbols). $R_H$ is weakly $T$-dependent below $T_{FL}$ and develops stronger $T$-dependence in the GCSM regime. $R_H$  varies roughly  linearly on  $\delta$  at all three temperatures shown in (\ref{subfig:nHA}).
  }
  \label{fig:DCcoeff}
\end{figure*}

We identify the GCFL and GCSM regimes, as well as the cross-over scale
$T_{FL}$ separating them, from the  T dependence  of the
longitudinal resistivity $\rho_{xx}$, shown in Fig.
(\ref{fig:rhoxx0}). {We identify the Fermi liquid temperature $T_{FL}$ using the resistivity, rather than the more conventional thermodynamic measures, such as heat capacity. The latter variables do actually give rather similar  values, but the resistivity seems most  appropriate for this study.  Our definition is that up to and below  $T_{FL}$, the resistivity  $\rho_{xx} \sim T^2$, while above
$T_{FL}$, $\rho_{xx}$ displays a more complex set of $T$ dependence as
outlined in Ref.~[\onlinecite{Ding2017}].}  The Fermi liquid
temperature has been quantitatively estimated in Ref. [\onlinecite{Perepelitsky2016}]:
\begin{equation}
\label{tfl}
T_{FL} \simeq 0.05 \times D \delta^{\alpha},
\end{equation}
{where $\delta$ is the hole density $\delta = 1 - n$. The exponent
$\alpha\sim 1.39$ within DMFT \cite{Rok}; this is the value we will use below. $\alpha$  is somewhat greater for ECFL within the scheme used in Ref. [\onlinecite{Perepelitsky2016}]} and  hence
$T_{FL}$ given by DMFT is slightly higher than that by ECFL, as can also be seen in Fig.~(\ref{fig:rhoxx0}). Consequently as  $n$ increases, the  the ECFL curves for  $\rho_{xx}$  lie above those from DMFT.

\subsection{Hall constant} 

\begin{figure*}\label{fig:HA}
  \centering
  \subfigure[]{\label{subfig:HAzT2}\includegraphics[width=.75 \columnwidth]{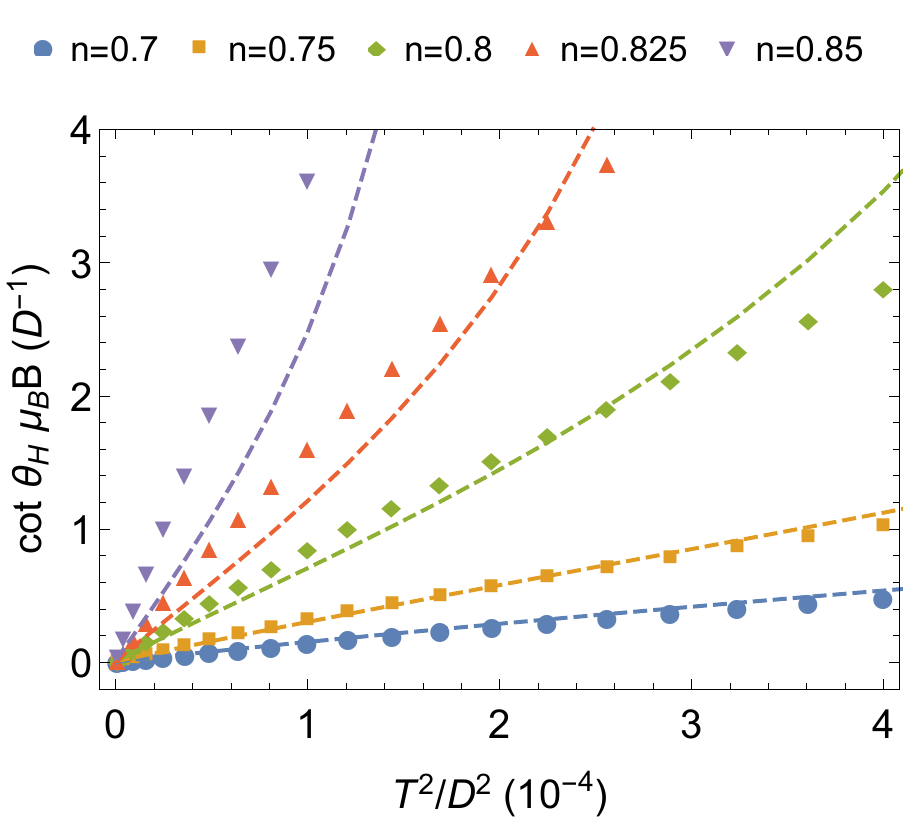}}
  \subfigure[]{\label{subfig:kink-ecfl-n07}\includegraphics[width=.75 \columnwidth]{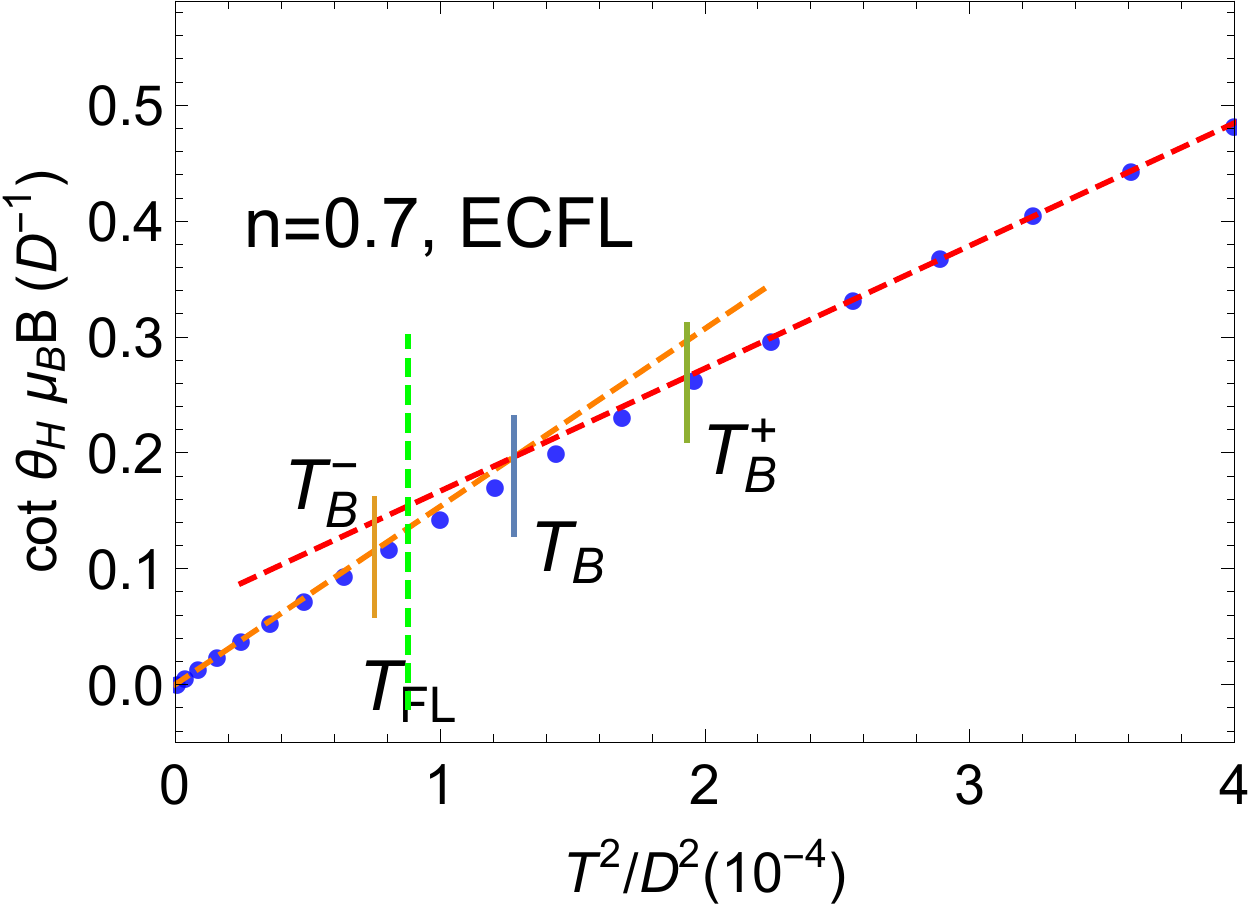}}
  \subfigure[]{\label{subfig:kink-dmft-n07}\includegraphics[width=.75 \columnwidth]{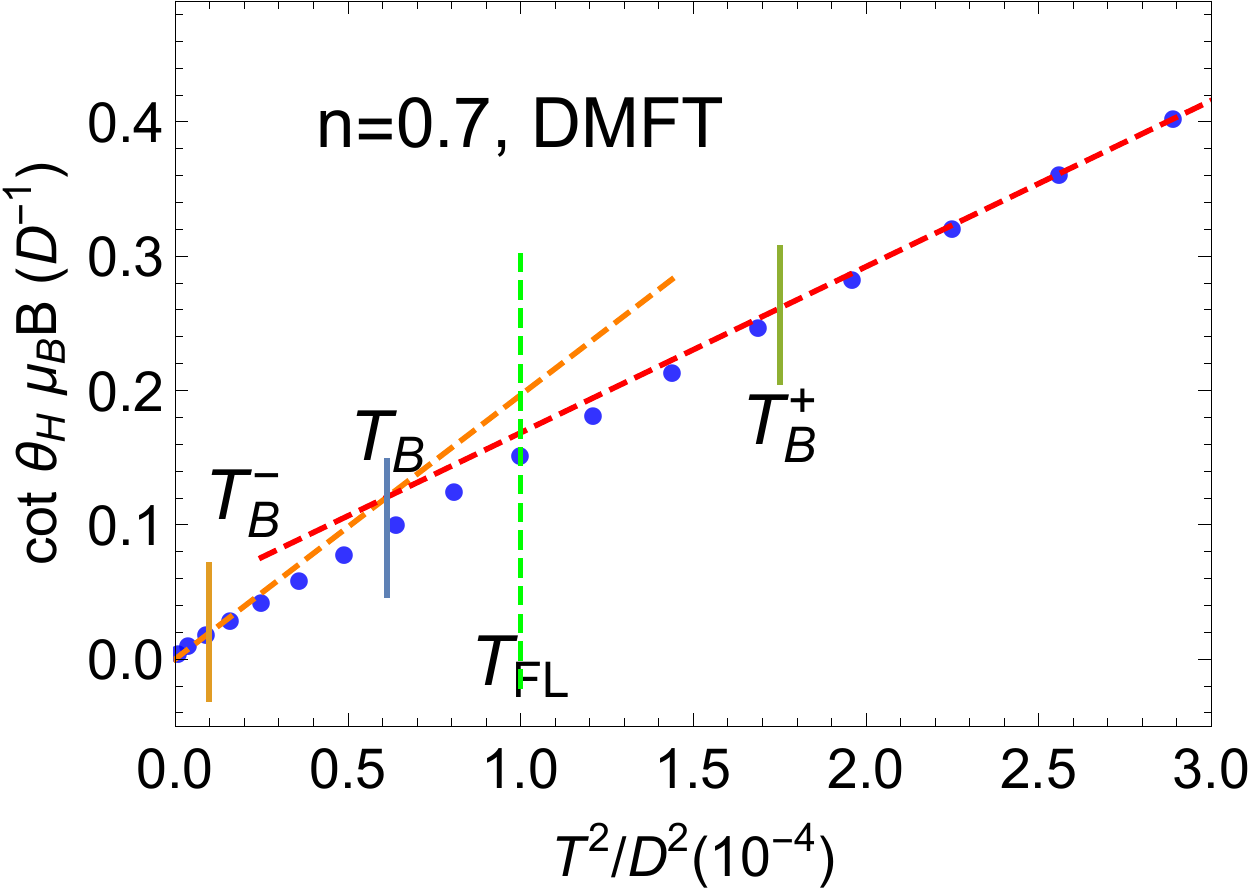}}
  \subfigure[]{\label{subfig:HA-Tbend}\includegraphics[width=.75 \columnwidth]{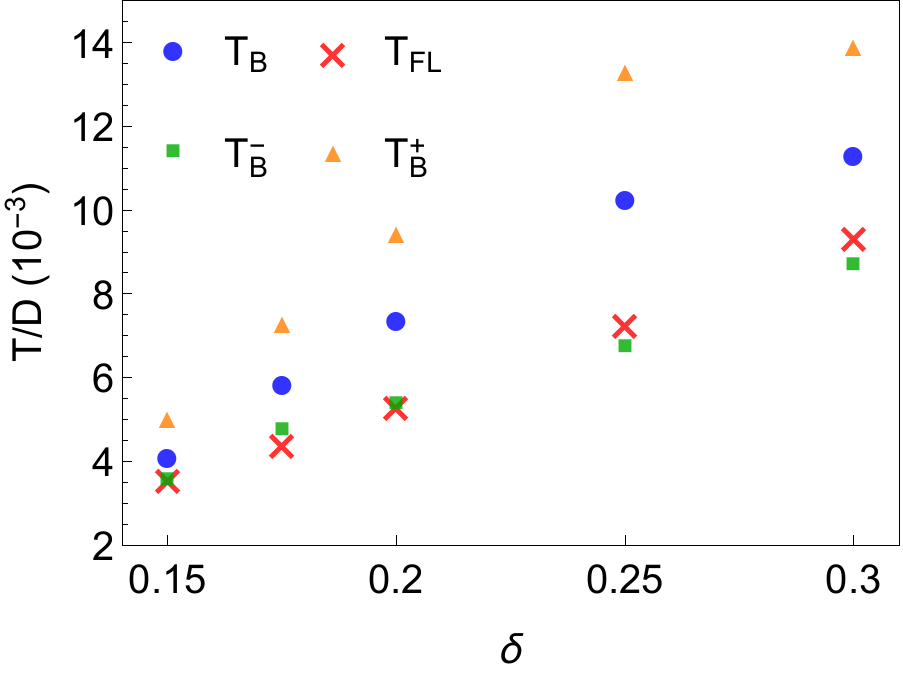}}
  \subfigure[]{\label{subfig:kink-ecfl}\includegraphics[width=.75 \columnwidth]{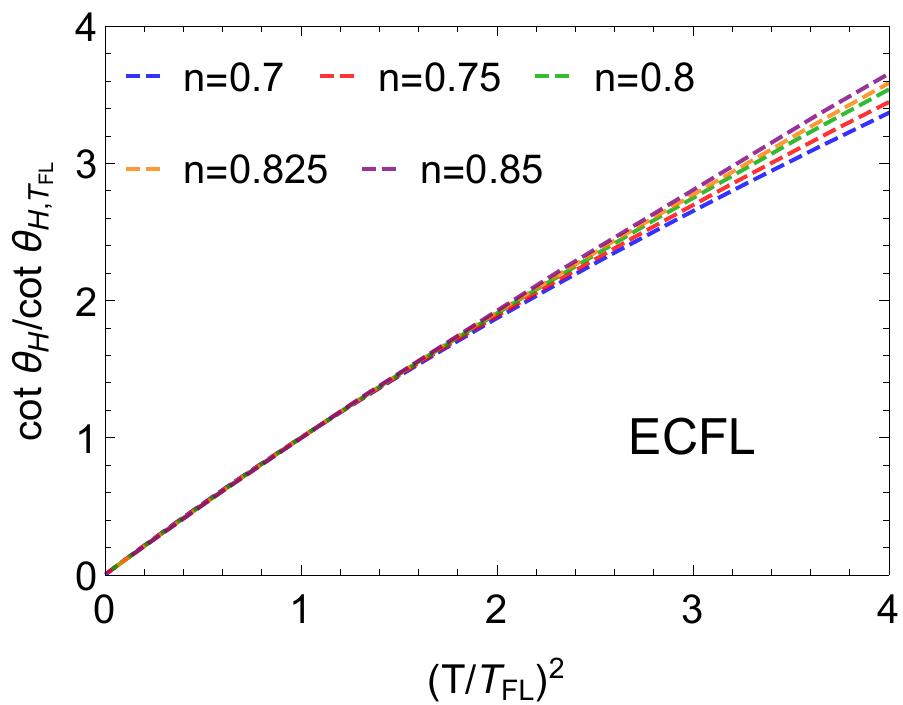}}
  \subfigure[]{\label{subfig:kink-dmft}\includegraphics[width=.75 \columnwidth]{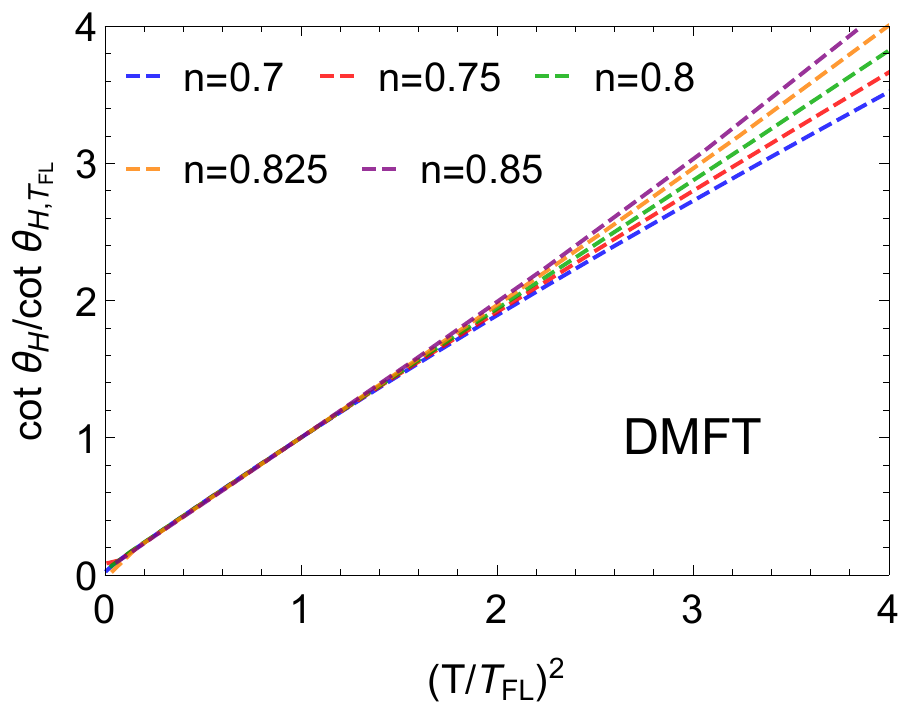}}
\caption{Temperature dependence of the cotangent Hall angle $\cot
    \theta_H B$ of both ECFL (symbols) and DMFT (dashed lines) shown
    as a function of $T^2$ (\ref{subfig:HAzT2}). The Hall angles $\cot
    \theta_H B \propto T^2$ in the GCFL regime, passes through a
    slight downward bend (i.e., a kink), and continues as $T^2$ within
    the temperature range studied. The downward bending regime is
    characterized by its onset $T_B^-$, the crossing of the two
    different $T^2$ lines $T_B$, and its ending $T_B^+$. {We
    illustrate the kink and the determination of $T_B$, $T_B^-$ and
    $T_B^+$ at $n = 0.7$ for both ECFL (\ref{subfig:kink-ecfl-n07})
    and DMFT (\ref{subfig:kink-dmft-n07}).  $T_B$, $T_B^-$ and $T_B^+$
    obtained from the ECFL are shown as a function of $\delta$ in (\ref{subfig:HA-Tbend}). We plot
    $\frac{\cot \theta_H }{\cot \theta_H(T=T_{FL})}$ as functions of
    $(T/T_{FL})^2$ for ECFL (\ref{subfig:kink-ecfl}) and DMFT
    (\ref{subfig:kink-dmft}) to show the systematic evolution of the
    kinks when the density varies.}
%Such bending is seen in experiments\cite{Chien1991,Hwang1994,Ando2004}, but the feature  appears unremarked upon.  From experiments, we find $T_B \simeq 100\ K, 80\ K, 70\ K$ for LSCO at $\delta = 0.21,\ 0.17,\ 0.14$ respectively from Fig. (3c) of Ref. [\onlinecite{Ando2004}]. The values are roughly comparable  with the theoretical results $T_B = 70\ K,\ 60\ K,\ 40\ K$ at $\delta = 0.2,\ 0.175,\ 0.15$ if we set $D = 10^4\ K$.
}
\end{figure*}

In Fig. (\ref{fig:DCcoeff}), we show $R_H$ as a function of
temperature at different densities for low temperatures $T<0.02 D$
(\ref{subfig:RH}), and as a function of the hole density $\delta =
(1-n)$ at $T= 0.002D, 0.005D, 0.01D$ (\ref{subfig:nHA}). The Hall
constant is weakly temperature-dependent for $T \ll T_{FL}$, but it
starts to decrease on warming, as seen in Fig.~(\ref{subfig:RH}).

As a function of density $\delta$ the  Hall constants from the two
theories  agree quite well, and are roughly linear with $\delta$. The
extrapolation to $\delta\to0$ is uncertain from the present data. One might be tempted to  speculate that it  vanishes, since the lattice density of states is particle-hole symmetric. This question deserves further study with different densities of states that break the particle-hole symmetry.

\subsection{ Cotangent of the   Hall  angle}

The theoretical  results for  cotangent of the  Hall angle $(\cot \theta_H) B = (\sigma_{xx}/\sigma_{xy}) B$
are shown as a function of $T^2$ in Fig.~(\ref{subfig:HAzT2}).   We see that {in DMFT as well as ECFL}, the $\cot(\theta_H)$ is linear in $T^2$ on both sides of a bend (or kink) temperature, which  increases with increasing hole density $\delta$. {However this kink is weaker in DMFT than in  ECFL.} This bending was already noted in Fig.~ (5.a) of Ref.~(\onlinecite{Shastry-Mai}), within the 2-d ECFL theory.
We may thus infer that $\cot(\theta_H)$ goes as $Q_{FL} T^2$ in the
Fermi liquid regime, passes through a slight downward bend, and
continues as $Q_{SM}T^2$ in the strange metal regimes, such that $Q_{FL} > Q_{SM}$.
 The difference, $A_{FL} - A_{SM}$,  becomes smaller as $\delta$ decreases.

 In order to characterize this kink more precisely, we define the
 downward bending regime by its onset temperature $T_B^-$, the
 crossing temperature of the two different $T^2$ lines $T_B$, and
 its ending temperature $T_B^+$. The temperatures $T_B^{-(+)}$ are
 determined by 5\% deviation from the $T^2$-fitting well below (above)
 $T_{FL}$, and $T_B$ is well defined as the crossing point of the two
 $T^2$-fittings. {We illustrate the kink and the determination of
 $T_B$, $T_B^-$ and $T_B^+$ at $n = 0.7$ for both ECFL in
 Fig.~(\ref{subfig:kink-ecfl-n07}) and DMFT in
 Fig.~(\ref{subfig:kink-dmft-n07}). } In Fig.~(\ref{subfig:HA-Tbend}),
 we show $T_B,\ T_B^-,\ T_B^+$ and $T_{FL}$ obtained from ECFL as
 functions of $\delta$. We see that $T_B^-$ is identical to $T_{FL}$,
 while $T_B$ and $T_B^+$ are $T_{FL}$ plus some constants with weak
 $\delta$-dependence. {We plot $\frac{\cot \theta_H }{\cot
 \theta_H(T=T_{FL})}$ as functions of $(T/T_{FL})^2$ for ECFL in Fig.
 (\ref{subfig:kink-ecfl}) and DMFT in Fig. (\ref{subfig:kink-dmft}) to
 show the systematic evolution of the kinks when the density is varied.}

\subsection{Kink in cotangent of the Hall angle}

There has been much interest in the quadratic $T$ dependence of
$\cot(\theta_H)$ in the literature \cite{Chien1991,Anderson1991}. It is
intriguing that a kink in the plot of $\cot(\theta_H)$ versus $T^2$
curves is seen in almost all experiments, although it appears
to not have been commented on earlier. Such bending is clearly seen in
experimental data Fig.~(2) of \refdisp{Chien1991}, Fig.~(4) of
\refdisp{Hwang1994} and Fig.~(3.c) of \refdisp{Ando2004}.

From Fig.~(3.c) of \refdisp{Ando2004}
we estimate $T_B \simeq 100\ K, 80\ K, 70\ K$ for LSCO at $\delta = 0.21,\ 0.17,\ 0.14$ respectively.
These are comparable with the ECFL results $T_B = 70\ K,\ 60\ K,\ 40\ K$ at $\delta = 0.2,\ 0.175,\ 0.15$, if we set $D = 10^4\ K$.
The trend of $T_B$ and the prefactor difference $A_{FL} - A_{SM}$ also
agrees with what we find, i.e., both $T_B$ and $A_{FL} - A_{SM}$ decrease as $\delta$ is lowered.
An increase of $A_{SM}$ at even higher temperatures is also observed in Ref.[\onlinecite{Ando2007}], similar as what we find in Fig. (\ref{subfig:HAzT2}) above the GCSM regime.

It is notable that the bending temperatures $T_B$ in theory and in
experiments are on a similar scale. It is therefore of interest to
explore this kink in $\cot(\theta_H)$ more carefully. From the
perspective of the ECFL and DMFT theories, we note that the kink
represents one of the basic crossovers discussed in
\refdisp{Ding2017}, namely from the GCFL to GCSM regimes. It would be
interesting to explore this feature more closely in experiments, in
particular to see if the theoretically expected correlation between
the crossover in $\rho_{xx}$ and $\cot(\theta_H)$ finds support.

\section{Optical response}
\label{sec:optical}

\subsection{Optical conductivity and the longitudinal scattering rates $\Gamma_{tr}$}\label{subsec:optical-conductivity}

\begin{figure*}[t]
  \centering
  \subfigure[]{\label{subfig:SigmaXXOmega1}\includegraphics[width=.85\columnwidth]{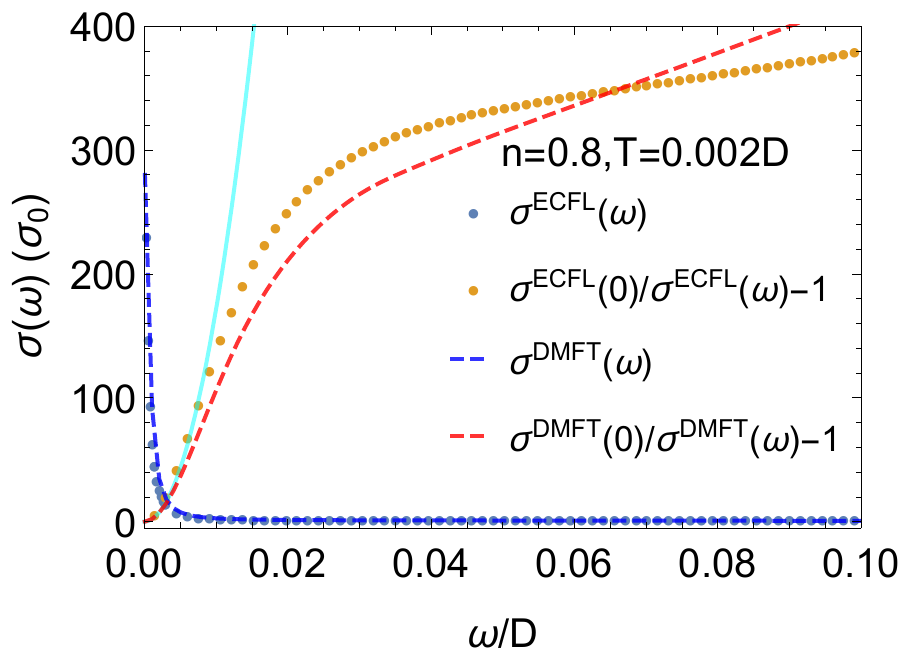}}
  \subfigure[]{\label{subfig:SigmaXXOmega2}\includegraphics[width=.85\columnwidth]{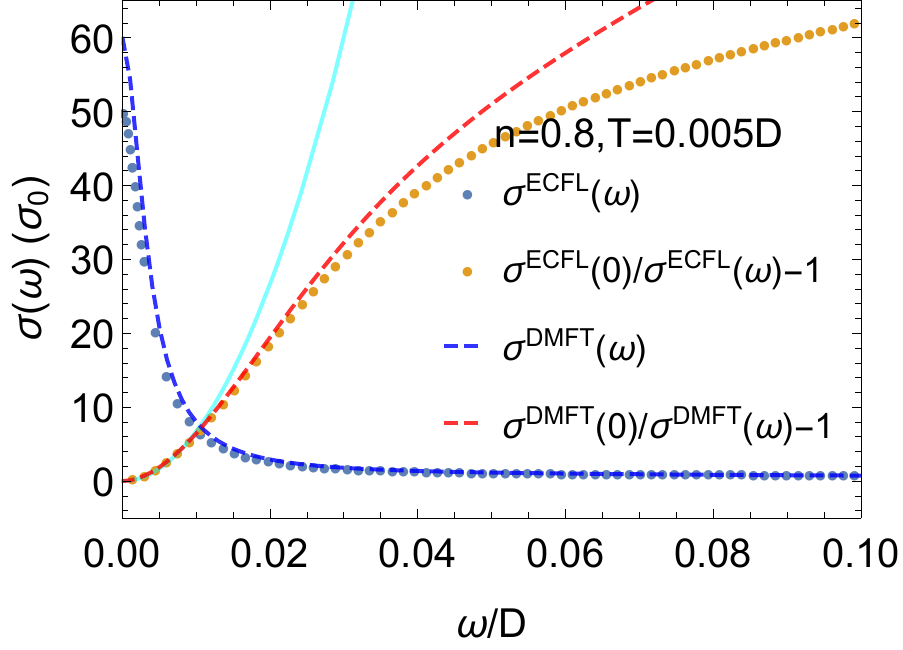}}
  \subfigure[]{\label{subfig:SigmaXXOmega3}\includegraphics[width=.85\columnwidth]{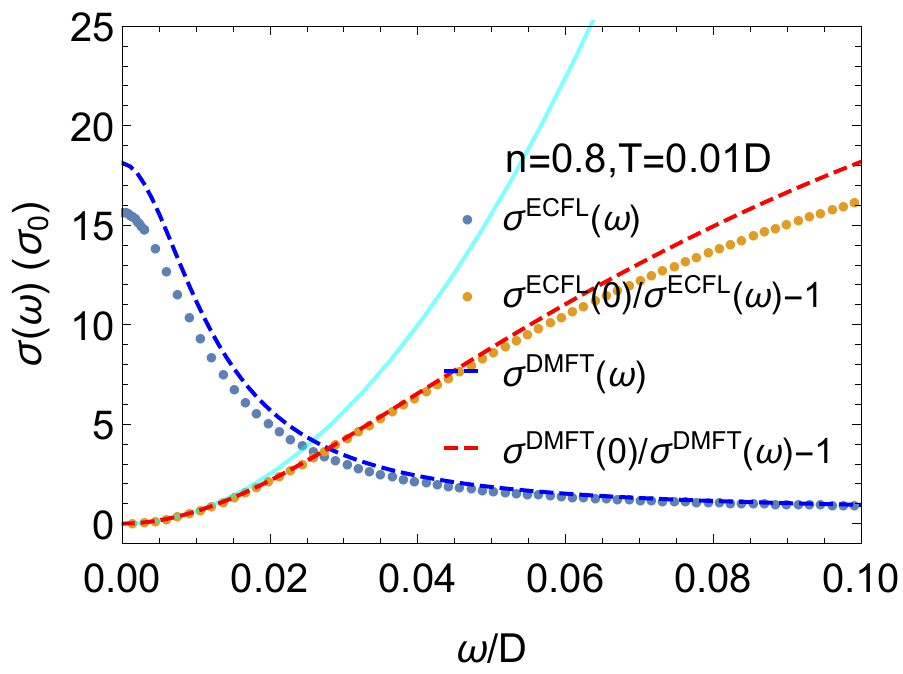}}
  \subfigure[]{\label{subfig:SigmaXXDrude}\includegraphics[width=.85\columnwidth]{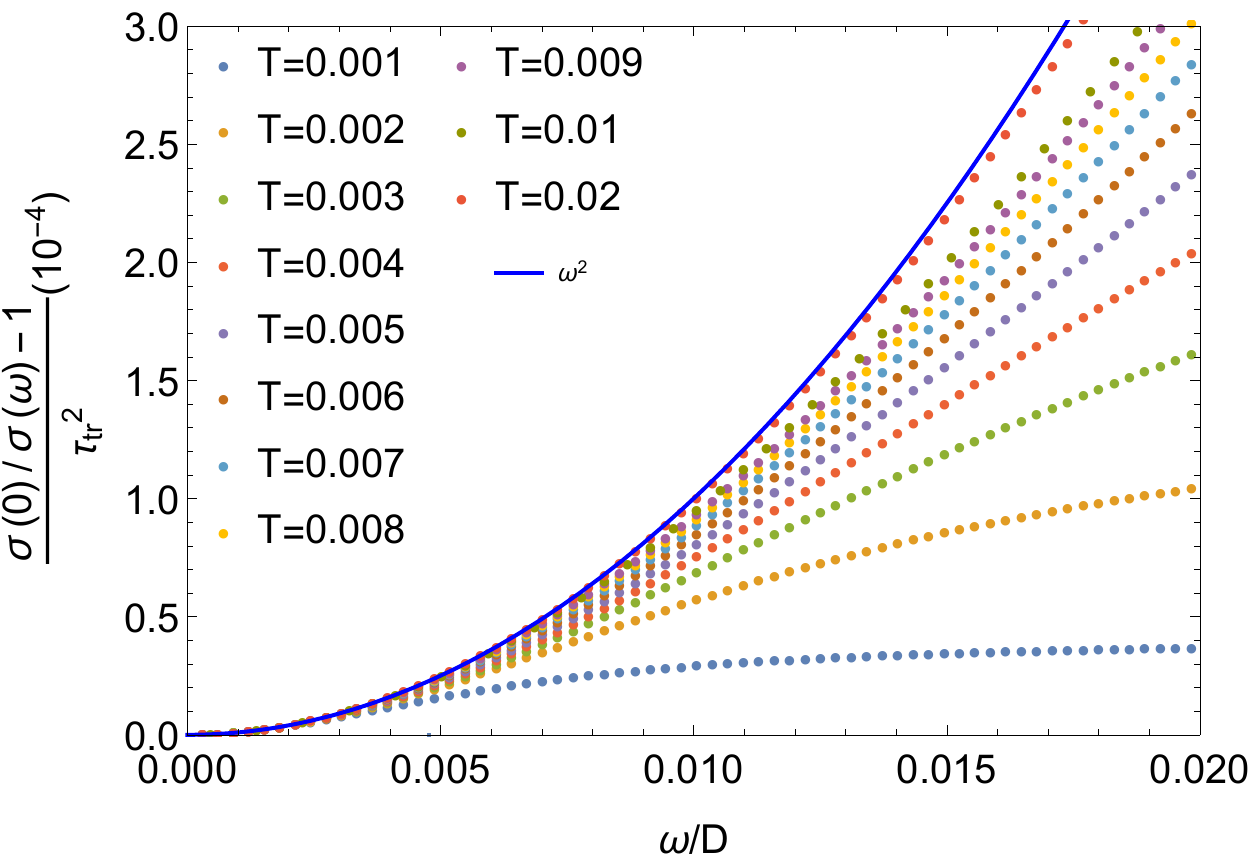}}
  %\subfigure[]{\label{subfig:SigmaXXDrudeWidth}\includegraphics[width=8cm]{./drude_width.pdf}}
  %\subfigure[]{\label{subfig:HallAngleOmega}\includegraphics[width=4cm]{./HallAngleOmega.pdf}}
  %
  \caption{$\sigma_{xx}(\omega)$ and $\sigma(0)/\sigma(\omega) - 1$
for $n=0.8$ at $T=0.002D$ (\ref{subfig:SigmaXXOmega1}), $T=0.005D$
(\ref{subfig:SigmaXXOmega2}) and $T=0.01D$
(\ref{subfig:SigmaXXOmega3}) for DMFT (dashed lines) and ECFL (solid
symbols). The cyan solid lines are $\omega^2$ fitting near $\omega \to
0$. In (\ref{subfig:SigmaXXDrude}) we normalize
$\sigma(0)/\sigma(\omega)-1$ curves computed from ECFL for various
temperatures by $\tau_{tr}^2$ with $\tau_{tr}$ obtained from the fits
at small frequencies to the Drude formula. The solid blue line is a 
$\omega^2$ curve. 
%As shown in (\ref{subfig:SigmaXXOmega1})
%(\ref{subfig:SigmaXXOmega2}) and (\ref{subfig:SigmaXXOmega3}), DMFT
%results agree well with ECFL within this range for all temperatures.
%Therefore, we only show ECFL results in (\ref{subfig:SigmaXXDrude}).
%Clearly identifiable  Drude peaks are found for low temperatures, which are almost delta function like. At high $T$  the $\omega^2$ regime broadens the peaks  become broader.
} \label{fig:sigma_xx_optical}
\end{figure*}

{In Fig. (\ref{fig:sigma_xx_optical}) we show the optical conductivity
$\sigma_{xx}(\omega)$ as well as the quantity
$\sigma_{xx}(0)/\sigma_{xx}(\omega) - 1$, which better presents the
approach to the zero frequency limit and is to be compared with the
Boltzmann RTA form (Drude formula) in Eq.~\eqref{eq:sigma-xx-tau}. We
display plots obtained from both ECFL (symbols) and DMFT (dashed
lines) for fixed $n=0.8$ and for three temperatures to show the
generic behavior at $T< T_{FL}$, $T\simeq T_{FL}$ and $T>T_{FL}$:
$T=0.002D$ (\ref{subfig:SigmaXXOmega1}), $T=0.005D$
(\ref{subfig:SigmaXXOmega2}) and $T=0.01D$
(\ref{subfig:SigmaXXOmega3}). {ECFL results agree well with the
exact solution of DMFT within this temperature range.}
  %DMFT results agree well with ECFL within this range for all temperatures.

$\sigma_{xx}(\omega)$ shows a narrow Drude peak below $T_{FL}$ which
broadens as $T$ increases and finally takes a form well approximated
by a broad Lorentzian at $T=0.01$ D. Correspondingly,
$(\sigma_{xx}(0)/\sigma_{xx}(\omega) - 1)$ is quadratic in frequency
and can be fit to $\tau_{tr}^2 \omega^2$ to extract the relaxation
time $\tau_{tr}$. The $\omega^2$ regime has a width $\propto
\tau^{-1}_{tr}$. The fitting is performed at very small frequencies
well within this quadratic regime. At higher frequency,
$(\sigma_{xx}(0)/\sigma_{xx}(\omega) - 1)$ flattens out and creates a
knee-like feature in-between. The flattening tendency decreases as $T$
increases, and $1/\sigma_{xx}(\omega)$ grows monotonically. This
knee-like feature thus becomes smoother as $T$ increases and
eventually is lost for $T > T_{FL}$. This trend is illustrated in
Fig.~(\ref{subfig:SigmaXXDrude}), where we normalize all curves of
$(\sigma_{xx}(0)/\sigma_{xx}(\omega) - 1)$ by their corresponding
$\tau_{tr}^2$, while the $\omega^2$ curve is shown as a solid blue
line. All curves fall onto the $\omega^2$ line at small frequencies,
and peal off at a frequency which increases as $T$ increases.

These scattering rates are shown as a function of temperature in Fig.~(\ref{subfig:gamma-tr}).
{The scattering rate $\Gamma$ has a similar temperature dependence
as the resistivity, i.e., a quadratic-T regime at low temperatures followed by a linear-T regime at higher temperatures.}
%and they agree with the $dc$ resistivity.
 }

\subsection{Optical Hall angle and the transverse scattering rates $\Gamma_{H}$}
%\begin{itemize}
%\item Drude peak at {\it low} temperatures
%\item $\Gamma_{H}$ can extracted within the Drude peak
%\item width of the Drude peak? more likely fixed, not increasing with $T$ or $\Gamma_H$
%\item non-Drude behavior $\sim \omega^a$ with increasing $a$ as T increases (analogous to DC)
%\end{itemize}

\begin{figure*}[t]
  \centering
  \subfigure[]{\label{subfig:thetaH-n08T002}\includegraphics[width=.95\columnwidth]{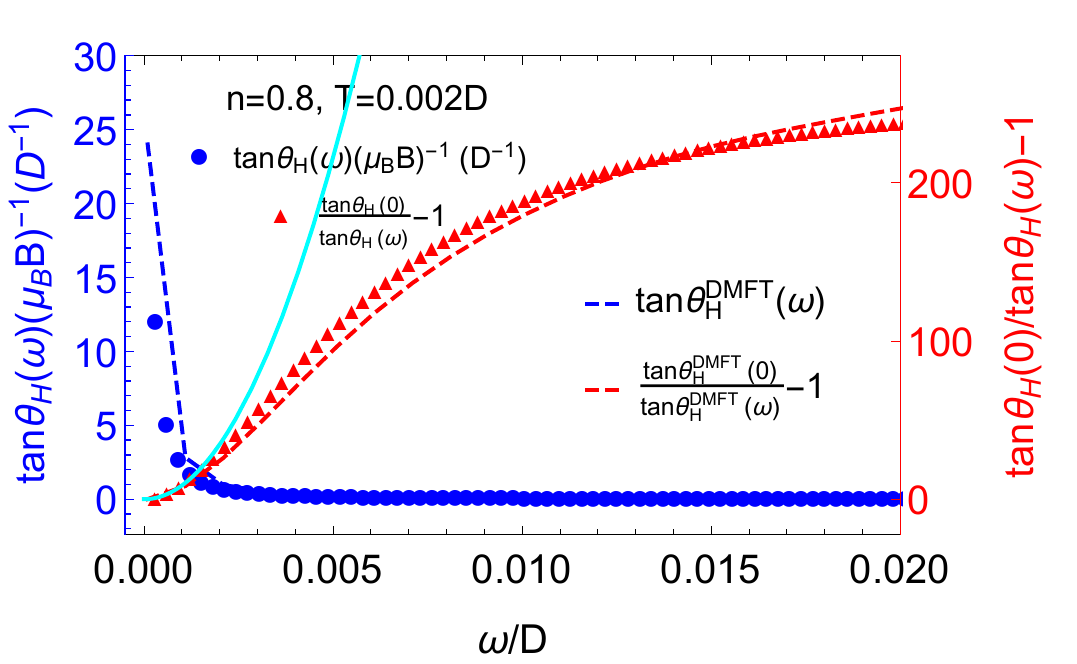}}
  \subfigure[]{\label{subfig:thetaH-n08T005}\includegraphics[width=.95\columnwidth]{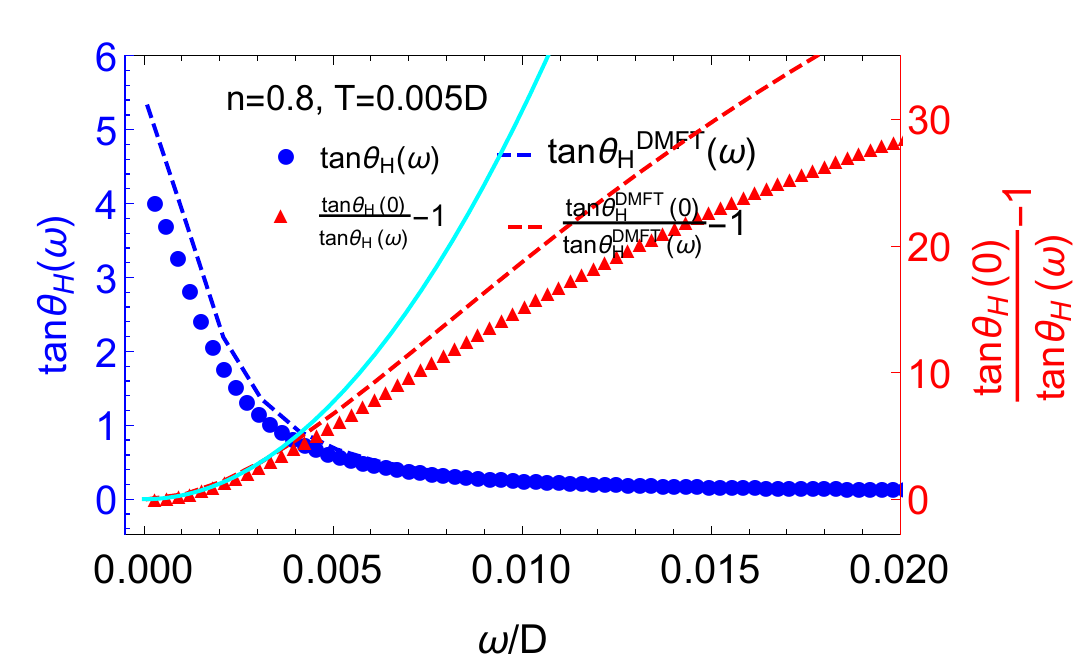}}
 \subfigure[]{\label{subfig:thetaH-n08T010}\includegraphics[width=.95\columnwidth]{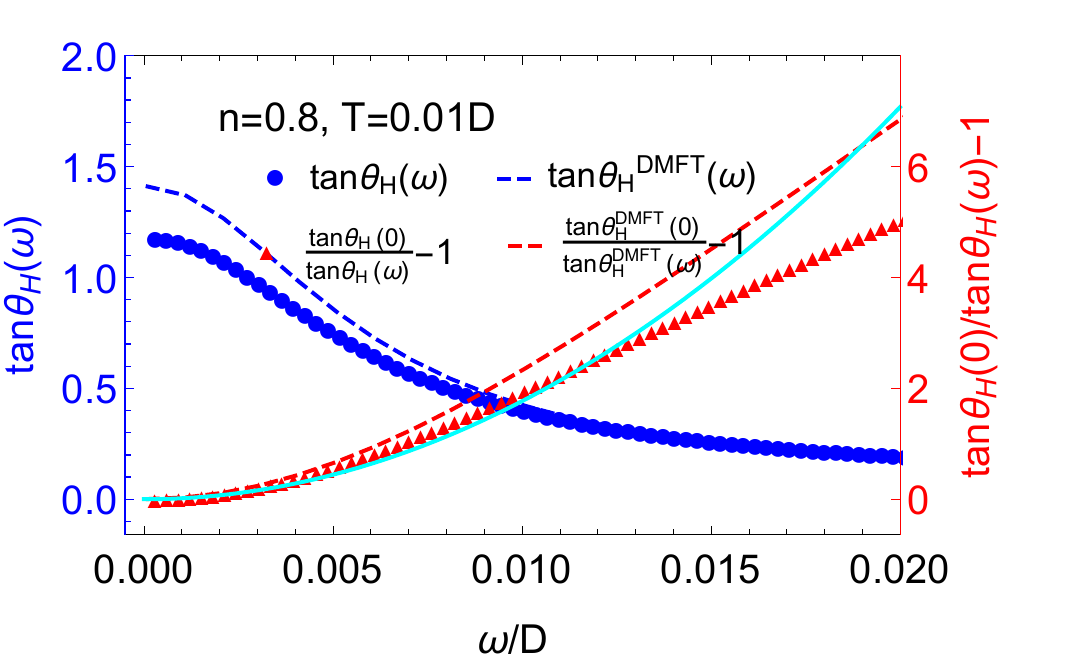}}
  \subfigure[]{\label{subfig:thetaH-drude-n07}\includegraphics[width=.85\columnwidth]{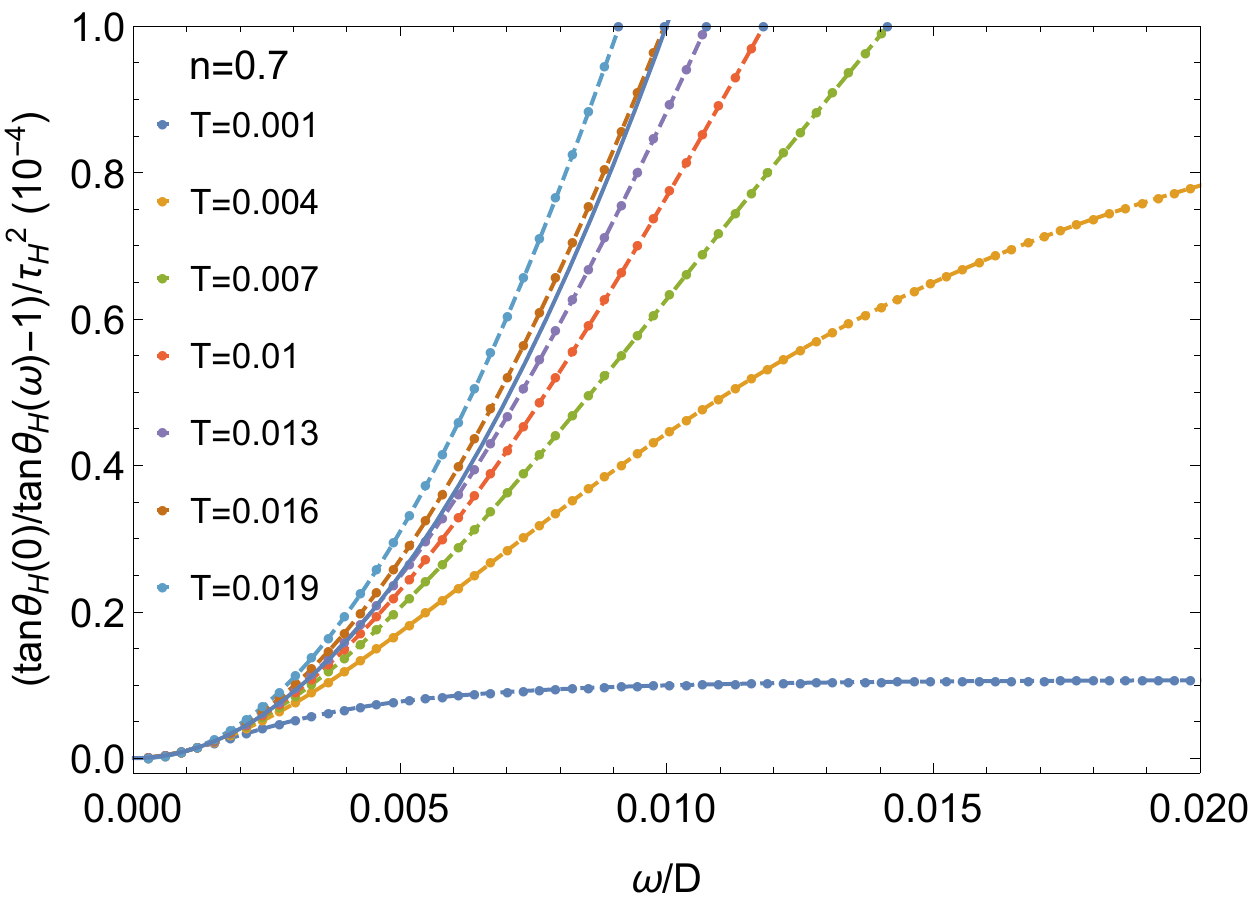}}
  \subfigure[]{\label{subfig:thetaH-drude-n08}\includegraphics[width=.85\columnwidth]{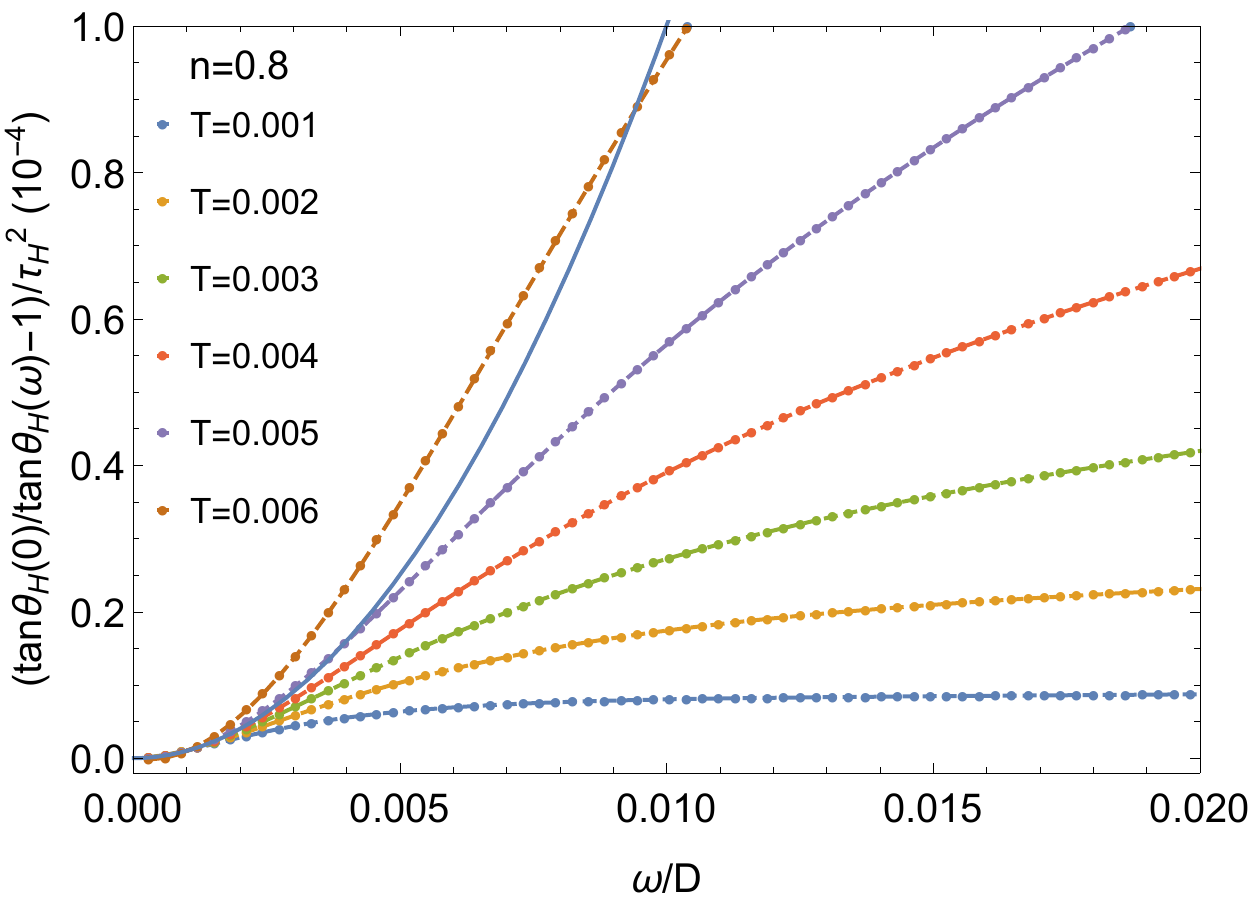}}
  \caption{Optical Hall angles $\tan\theta_H(\omega)$ (blue) and $\tan\theta_H(0)/\tan\theta_H(\omega)-1$(red) shown for $n=0.8$, $T=0.002D$ (\ref{subfig:thetaH-n08T002}), $T=0.005D$ (\ref{subfig:thetaH-n08T005}) and $T=0.01D$ (\ref{subfig:thetaH-n08T010}) for DMFT (dashed lines) and ECFL (solid symbols). The cyan solid lines are $\omega^2$ fitting near $\omega \to 0$.
    $[\tan\theta_H(0)/\tan\theta_H(\omega)-1]/\tau_H^2$ obtained from ECFL shown for $n=0.7$ (\ref{subfig:thetaH-drude-n07}) and $n=0.8$ (\ref{subfig:thetaH-drude-n08}).
Drude peaks are found to be narrow (note the different horizontal axis
scale compared to Fig.~3).% and becomes narrower as $T$ increases or when doping level $\delta$ decreases.
  }
  \label{fig:theta-H}
\end{figure*}

In Fig.~(\ref{fig:theta-H}), we show the optical tangent Hall angle
$\tan \theta_H(\omega)$ and the quantity $\tan \theta_H(0) / \tan
\theta_H(\omega) -1$. We display plots obtained from both ECFL
(symbols) and DMFT (dashed lines) for fixed $n=0.8$ and for three
temperatures to show the generic behavior at $T< T_{FL}$, $T\simeq
T_{FL}$ and $T>T_{FL}$: $T=0.002D$ (\ref{subfig:thetaH-n08T002}),
$T=0.005D$ (\ref{subfig:thetaH-n08T005}) and $T=0.01D$
(\ref{subfig:thetaH-n08T010}). The ECFL results agree well with those
from DMFT within this temperature range.

Just like $\sigma_{xx}(\omega)$, $\tan\theta_H(\omega)$ possesses a
narrow Drude peak below $T_{FL}$ that broadens in a similar way with
increasing temperature. $(\tan \theta_H(0) / \tan \theta_H(\omega)
-1)$ is quadratic in frequency and we fit $\tau_H^2 \omega^2$ to
extract the transverse relaxation time $\tau_H$. The $\omega^2$
regime, however, has a very narrow, weakly $T$-dependent width which
is about $0.003\ D$. The relaxation time $\tau_H$ is extracted by
fitting within this very low frequency range. Above this energy a
flattening behavior, similar to that in the optical conductivity,
takes place at low temperatures. At higher temperatures and lower hole
density, a power-law behavior with an exponent that increases with $T$
gradually replaces the flattening out behavior. Such a tendency is
visible in Figs. (\ref{subfig:thetaH-drude-n07}) and
(\ref{subfig:thetaH-drude-n08}), where all $(\tan \theta_H(0) / \tan
\theta_H(\omega) -1)$ curves are normalized by their corresponding
$\tau_H^2$.

In Fig. (\ref{subfig:gamma-H}) we show $\Gamma_H$ (defined as
$\Gamma_H \equiv \tau_H^{-1}$) for various densities and temperatures
obtained from the Drude formula fitting. Their $T$-dependence
{is quadratic for both GCFL and GCSM regimes.}
%, similar as the $dc$ cotangent Hall angle.

\subsection{Emergence of two relaxation times}

\begin{figure*}[t]
  \centering
   \subfigure[]{\label{subfig:gamma-tr}\includegraphics[width=.85\columnwidth]{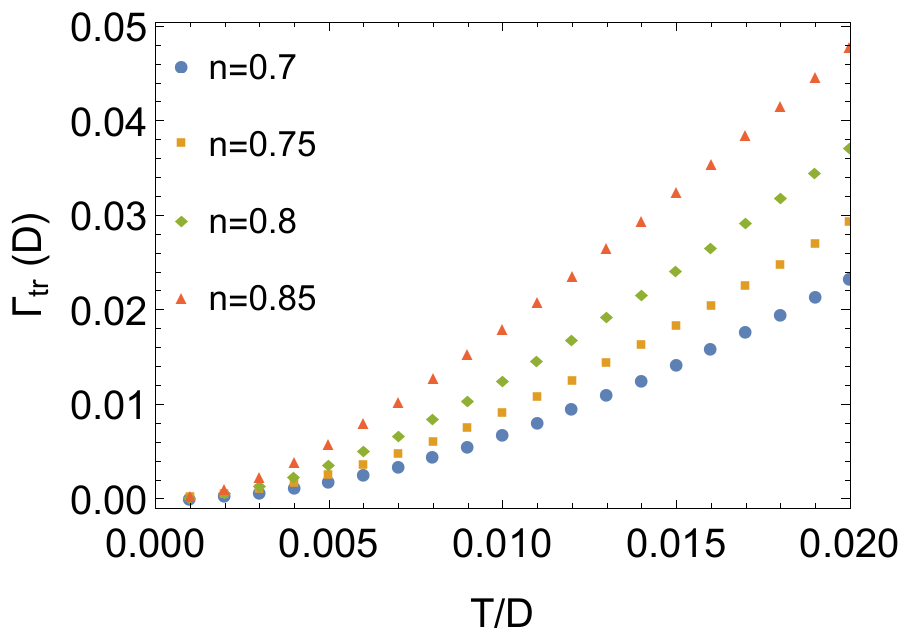}}
   \subfigure[]{\label{subfig:gamma-H}\includegraphics[width=.85\columnwidth]{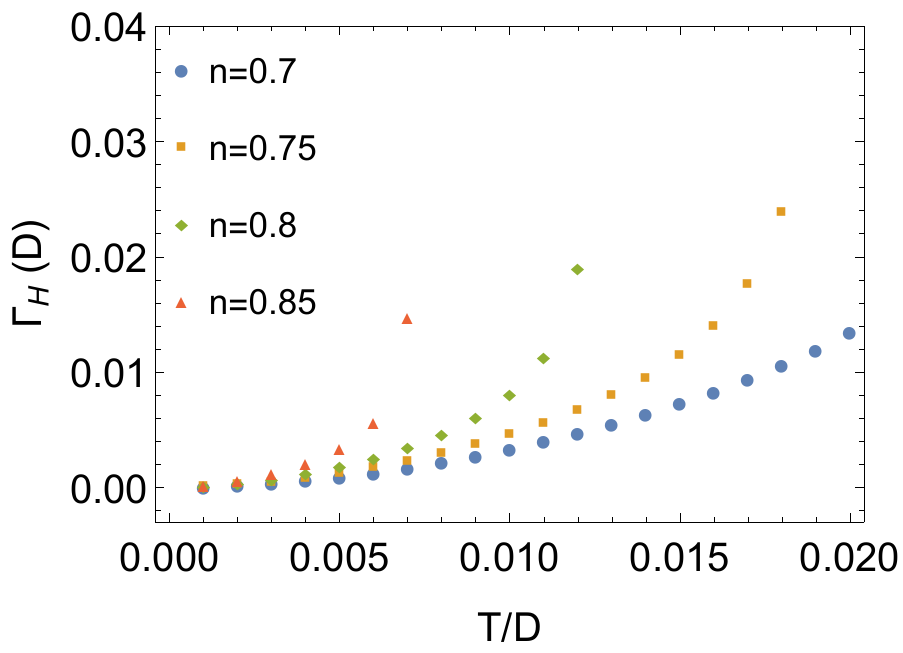}}
   \subfigure[]{\label{subfig:gamma-ratio} \includegraphics[width=.85\columnwidth]{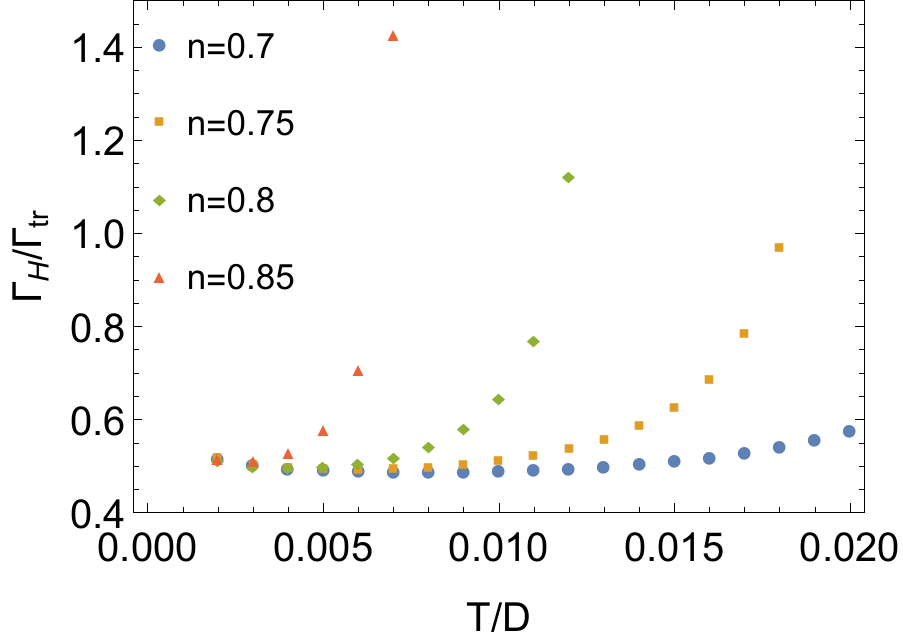}}
   \subfigure[]{\label{subfig:gamma-ratio-resc} \includegraphics[width=.85\columnwidth]{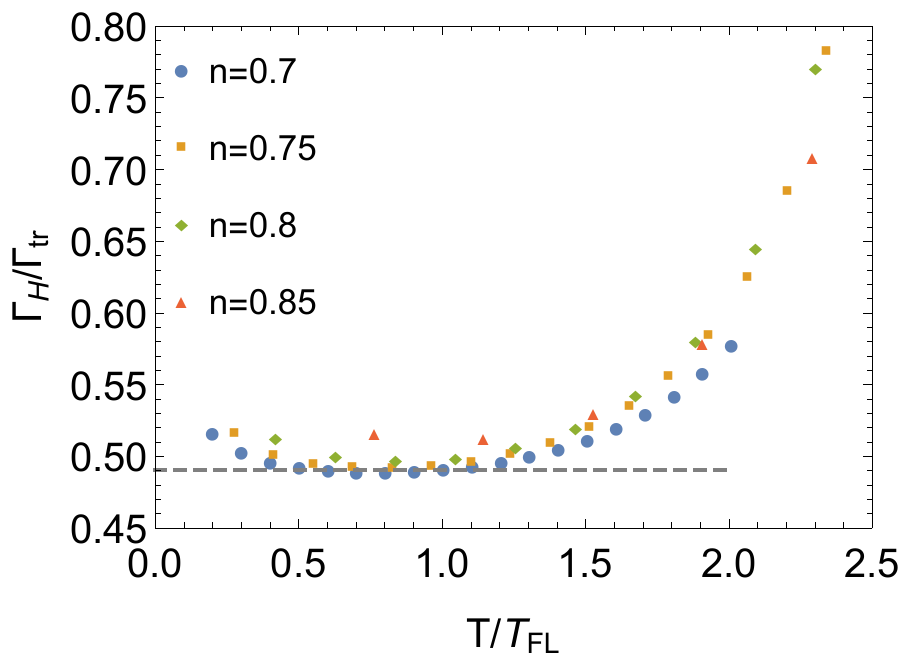}}
   %\subfigure[]{\label{subfig:srr-H}\includegraphics[width=8cm]{./GammaH_GammaTr.pdf}}
  %\subfigure[]{\label{subfig:srr-H-ht}\includegraphics[width=8cm]{./GammaH_GammaTr_HighT.pdf}}
  %\subfigure[]{\label{subfig:srr-theta}}
  %\subfigure[]{\label{subfig:srr-theta-ht}\includegraphics[width=8cm]{./figThetaTrZ.pdf}}
  %\subfigure[]{\label{subfig:srr-theta-H}\includegraphics[width=8cm]{./GammaTheta_GammaH2.pdf}}./GammaTheta_GammaTr_HighT.pdf
  %\subfigure[]{\label{subfig:srr-theta-H-ht}\includegraphics[width=8cm]{./GammaTheta_GammaH_HighT2.pdf}}
  %
   \caption{Longitudinal relaxation rate $\Gamma_{tr}$ extracted by
   fitting $\sigma_{xx}(\omega)$ by the Drude formula
   (\ref{subfig:gamma-tr}), transverse relaxation time $\Gamma_H$
   extracted from $\theta_H(\omega)$ (\ref{subfig:gamma-H}), their
   ratio $\Gamma_{H}/\Gamma_{tr}$ as functions of $T$
   (\ref{subfig:gamma-ratio}) and as functions of scaled temperature
   $T/T_{FL}$ (\ref{subfig:gamma-ratio-resc}). All the relaxation
   rates are extracted from the ECFL optical response results. 
   %Both
   %scattering rates agree qualitatively with the $dc$ transport
   %coefficients. The $\Gamma_{H}/\Gamma_{tr}$ plotted as $T/T_{FL}$
   %clearly reveals the onset of the second {separate} scattering
   %rate at $T = T_{FL}$. 
   } \label{fig:Scattering-Rates-Ratios}
\end{figure*}

In Fig. (\ref{subfig:gamma-ratio}), we show $\Gamma_{H}/ \Gamma_{tr}$
as a function of temperature. At all densities considered this ratio
behaves differently for $T$ below and above $T_{FL}$. Below $T_{FL}$,
the ratio $\Gamma_{H}/\Gamma_{tr} \simeq 0.5$ remains essentially
constant, and hence the optical transport is dominated by a single
scattering rate. Once $T_{FL}$ is crossed, however,
$\Gamma_{H}/\Gamma_{tr}$ becomes strongly $T$-dependent. This 
indicates that 
%$\Gamma_{H}$ and $\Gamma_{tr}$ %become
%independent,e.g.
there are {\it two relaxation times} in the GCSM regime. This is
possible since the quasiparticles are no longer well defined for
$T>T_{FL}$, and different frequency regimes present in the spectral
functions contribute differently to the two relaxation times. {In Fig.
(\ref{subfig:gamma-ratio-resc}), we plot $\Gamma_{H}/ \Gamma_{tr}$
versus the rescaled temperature $T/T_{FL}$ to illustrate the clearly
distinct behavior below and above $T_{FL}$.}

\section{Analysis}\label{sec:analysis}

{ We begin by analyzing the exact formulas for the conductivities
$\sigma_{xx},\sigma_{xy}$ of Eqs. (\ref{eq:sxx-ep}) and
(\ref{eq:sxy-ep}), following \refdisp{voruganti} and
[\onlinecite{Perepelitsky2016}]} {within ECFL theory where more analytic insight is available.}

It has long been noted that the particle-hole asymmetry of the
spectral function is one of the characteristic features of strongly
correlated
systems\cite{renner1998,anderson2006,hanaguri2004,casey2008,pasupathy2008,gweon2011,Shastry2012}.
The dynamic particle-hole transformation is
defined by simultaneously inverting the wave vector and energy in
$\rho_G(\k, \omega)$ relative to the chemical potential $\bm{\mu}$ as
$ (\bm{\hat{k}}, \omega) \rightarrow - (\bm{\hat{k}}, \omega)$, with
$\bm{\hat{k}} = \k - \k_F$\cite{Shastry2012}. In the limit of $d \to
\infty$, we ignore the $\bm{\hat{k}}$ part of the transformation.
Consequently, the dynamic particle-hole asymmetry solely stems from
the asymmetry of the self-energy spectral function
$\rho_\Sigma(\omega, T) = - \Imm \Sigma(\omega,T)/\pi$. Instead of
analyzing $\rho_G$, we can simply focus on $\rho_{\Sigma}$ since
\begin{eqnarray}
  \rho_G &=& \frac{\rho_\Sigma}{ (\omega +\mu - \epsilon -\Ree\Sigma)^2 + \pi^2 \rho_\Sigma^2}.
  \label{eq:rhoG} \\
  &=& \frac{1}{\pi} \frac{B(\omega,T)}{(A(\omega,T)-\epsilon)^2+B^2(\omega,T)}
\end{eqnarray}
where
\begin{eqnarray}
  A(\omega,T) = \omega + \mu - \Ree  \Sigma(\omega,T), \\
  B(\omega, T) = \pi \rho_\Sigma(\omega, T) = - \Imm \Sigma(\omega,T).
\end{eqnarray}

Then we approximate the exact equations (\ref{eq:sxx-ep}) and
(\ref{eq:sxy-ep}) by their asymptotic values at low enough T,
following Ref. [\onlinecite{Perepelitsky2016}].  The idea is to first
integrate over the band  energy $\epsilon$ viewing one of the powers
of $\rho_G$ as  a $\delta$ function constraining $\epsilon\to A(\omega,T)$. This gives
\begin{eqnarray}
  & \sigma_{xx} = \frac{\sigma_0 D}{\Phi^{xx}[0]}  \int d\omega (-f') \frac{\Phi^{xx}[A(\omega,T)]}{B(\omega,T)},\label{eq:sigma-xx-asymp}\\
  &  \sigma_{xy} = \frac{\sigma_0 D q_e}{\Phi^{xx}[0]}  \int d\omega (f') \Big(\frac{\pt_\omega^2 \Phi^{xy}[A(\omega,T)]}{3 } + \frac{\Phi^{xy}[A(\omega,T)]}{2 (B(\omega,T))^2} \Big), \nonumber \\
  \label{eq:sigma-xy-asymp}
\end{eqnarray}
The first term in Eq.~(\ref{eq:sigma-xy-asymp}) turns out to be negligible compared to the second, and hence we will ignore it.
Next, we track down the electronic properties that give rise to a second relaxation time using the above asymptotic expressions.

{
\begin{figure}
  \centering
  \includegraphics[width=.85\columnwidth]{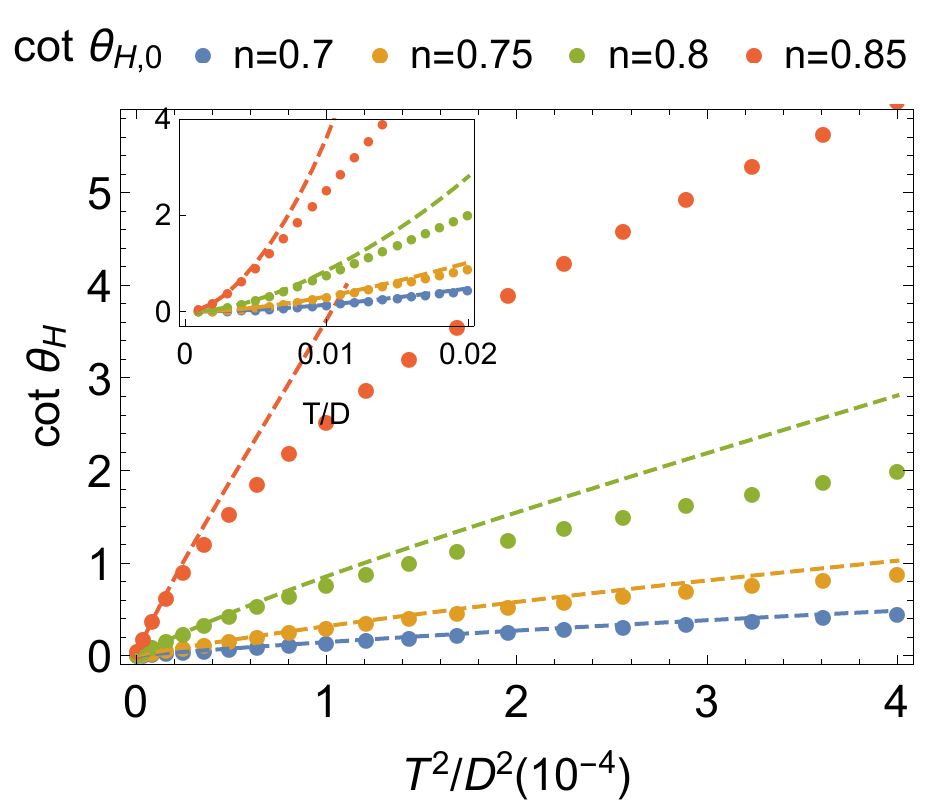}
     \caption{Zeroth order asymptotic cotangent Hall angles $\cot \theta_{H,0}$ plotted as functions of $T^2$ (main panel, symbols) compared with the exact results (dashed lines) and as functions of $T$ (inset).}
   \label{fig:cot-theta-H0}
 \end{figure}
  To the lowest order of approximation at low temperatures, we can make the substitution $f'(\omega) \to - \delta(\omega)$ in Eq.~(\ref{eq:sigma-xx-asymp})  and (\ref{eq:sigma-xy-asymp}), which gives
  \begin{equation}
    \label{eq:cot-theta0}
    \cot \theta_{H,0} / B =  \frac{2 B(0,T)}{q_e A(0,T)}.
  \end{equation}
  We show $\cot \theta_{H,0}$ in Fig. (\ref{fig:cot-theta-H0}). When
  plotted as a function of $T^2$ as shown in the main panel of
  Fig.~(\ref{fig:cot-theta-H0}), $\cot \theta_{H,0}$ (solid symbols) is in good agreement with the exact results (dashed lines) both qualitatively, i.e., showing a kink-like feature, and quantitatively except for relatively high temperatures and densities. However, when it is plotted as a function of $T$ (inset of Fig.~(\ref{fig:cot-theta-H0})), we find that the "kink" is actually the crossover from a $T^2$ behavior to a linear-$T$ behavior and $\cot \theta_{H,0}$ follows the $T$-dependence of $\rho_{xx}$. The lowest order approximation is insufficient to capture and to understand the second $T^2$ regime.
Therefore, we pursue more accurate asymptotic expressions of Eqs. (\ref{eq:sigma-xx-asymp}) and (\ref{eq:sigma-xy-asymp}).
}
%In Eq. (\ref{eq:sigma-xx-asymp}) and (\ref{eq:sigma-xy-asymp}), the anti-symmetric component of $\rho_{\Sigma}$ contribute in two ways. Firstly, it  leads to $\Ree  \Sigma(\omega = 0,T) \neq 0$ which contributes to $\Phi^{xx}[A(\omega = 0,T)]$.
%However, $A(0,T)$ hence $\Phi^{xx}[A(0,T)]$ has been shown to be small and almost independent of $T$ in \refdisp{Perepelitsky2016} and [\onlinecite{Ding2017}].  The most important contribution comes from $1/B(\omega,T)$.
Following Ref. [\onlinecite{Xu2013d}] and [\onlinecite{Ding2017}], we do the following small frequency expansion:
\begin{align}
  \label{eq:A-B-expansion}
  \begin{split}
    & \Phi^{xx (xy)} [A(\omega, T)] =  \Phi^{xx (xy)} [A_0]  \\
    & + \Phi^{xx (xy)\prime} [A_0] A_1\ \omega + \dots,
\end{split}
  \\
    & B(\omega, T) = B_0 + B_1 \omega + B_2 \omega^2 +  \dots, %B_3 \omega^3 +
\end{align}
where $A_0$ and $A_1$ is given by the expansion
\begin{equation}
A(\omega, T) = A_0 + A_1 \omega + \dots,
\end{equation}
Recall that $A_1 = Z^{-1}$, it is therefore large. In order to provide
further context to these coefficients $B_n$ and to connect with earlier discussions of the self energy, it is useful to recall a useful and suggestive expression for the imaginary self energy exhibiting particle-hole asymmetry at $k_F$ at low $\omega$ (e.g. see Eq.~(28) in \refdisp{Rok})
\begin{equation}
- \Im \, m \Sigma(\omega,T) \sim \pi \frac{(\omega^2 + \pi^2
T^2)}{\Omega_\Sigma(T)} \left(1- \frac{\omega}{\Delta}\right),
\end{equation}
where $\Omega_\Sigma$   behaves as $\sim Z^2$ in the low-$T$ Fermi
liquid regime. The scale $\Delta$ breaks the particle-hole symmetry of the leading term.
% It is T independent in the low $T$ Fermi liquid regime and has been discussed earlier.

{The variation of $\Omega_\Sigma$  and $\Delta$ in the GCSM regime is illustrated below in Fig.~(\ref{fig:Bs}).}
 Expanding this expression at low $\omega$ we identify the
 coefficients $B_0= \pi \frac{\pi^2 T^2}{\Omega_\Sigma(T)}$, $B_1= -
 \frac{B_0}{\Delta}$, $B_2=\frac{\pi}{\Omega_\Sigma}${, all of which are numerically verified to be valid for all temperatures we study in this work}. The negative
 sign of $B_1$ is easily understood.
%From this expression we see that the widely held relations
%\begin{align}
%B_0 = \pi^2 T^2 B_2,\label{eq:b0b2}\\
%B_1 = \pi^2 T^2 B_3 \label{eq:b1b3}
%\end{align}
\begin{figure*}[t]
  \centering
   \subfigure[]{\label{subfig:Omega}\includegraphics[width=.65\columnwidth]{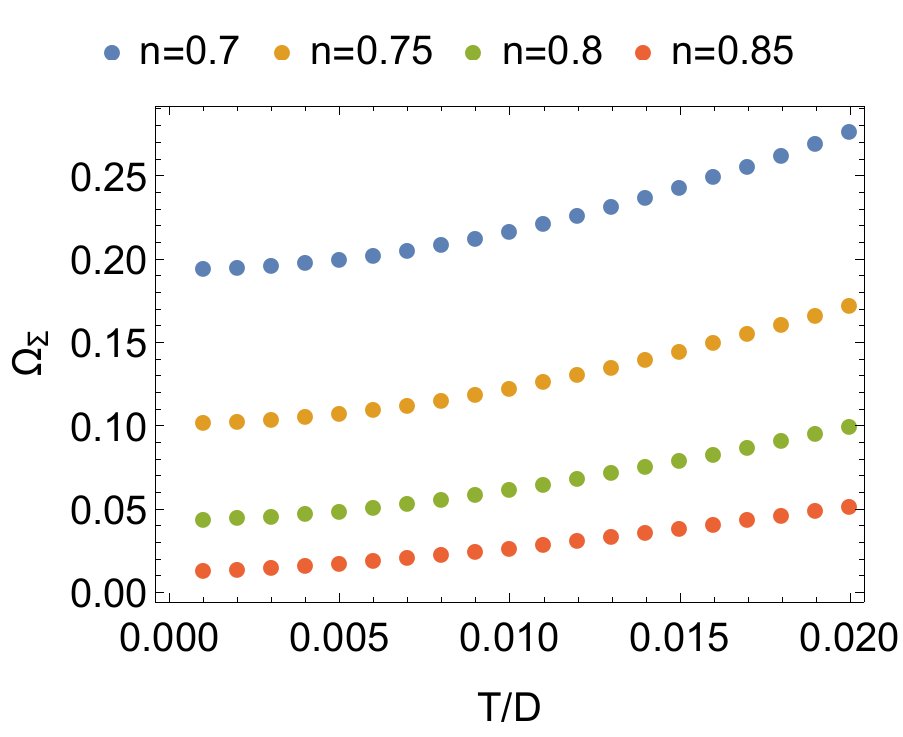}}
   \subfigure[]{\label{subfig:Delta}\includegraphics[width=.65\columnwidth]{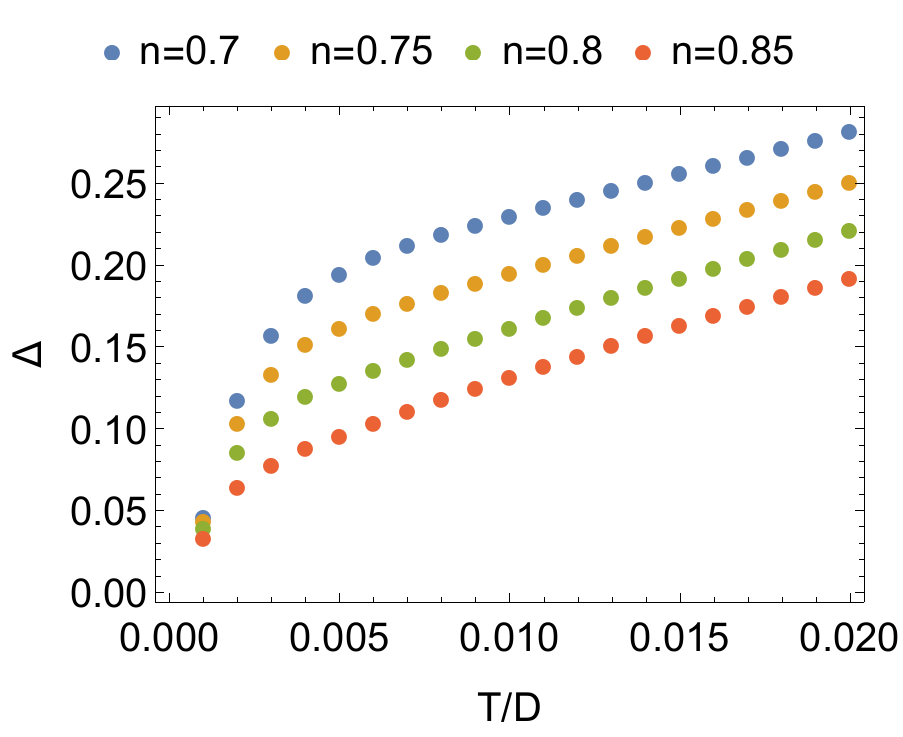}}
   %\subfigure[]{\label{subfig:B2}\includegraphics[width=.65\columnwidth]{./B2.pdf}}
   \caption{Coefficients of the small frequency expansion of the ECFL Dyson self-energy $\Omega_{\Sigma}$ (\ref{subfig:Omega}) and $\Delta$ (\ref{subfig:Delta}) plotted as functions of temperature.}
   \label{fig:Bs}
 \end{figure*}

\begin{figure*}[t]
  \centering
  \subfigure[]{\label{subfig:tauh1-tauh}\includegraphics[width=.65\columnwidth]{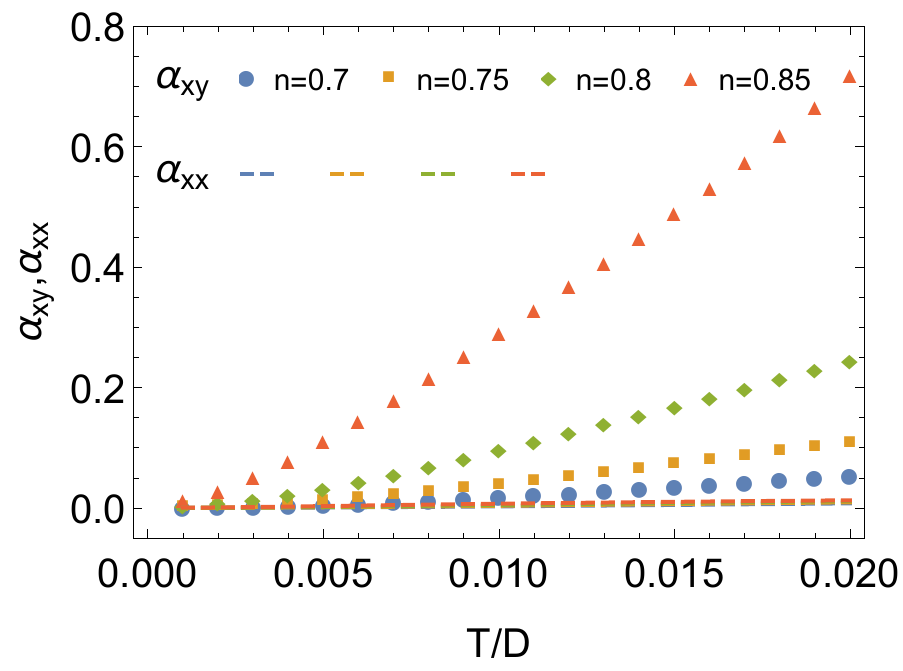}}
   \subfigure[]{\label{subfig:rhoxx-star}\includegraphics[width=.65\columnwidth]{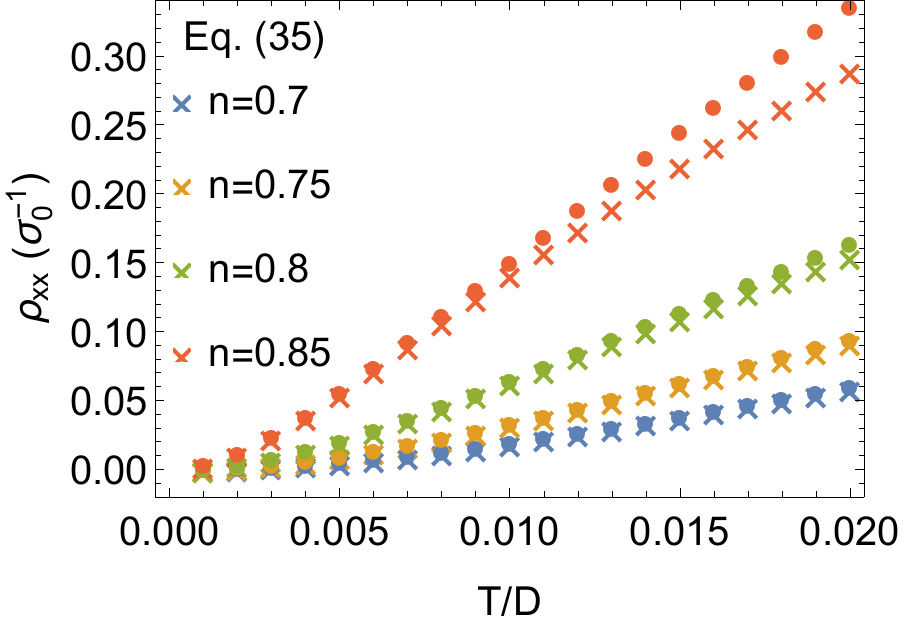}}
   \subfigure[]{\label{subfig:HA-star}\includegraphics[width=.65\columnwidth]{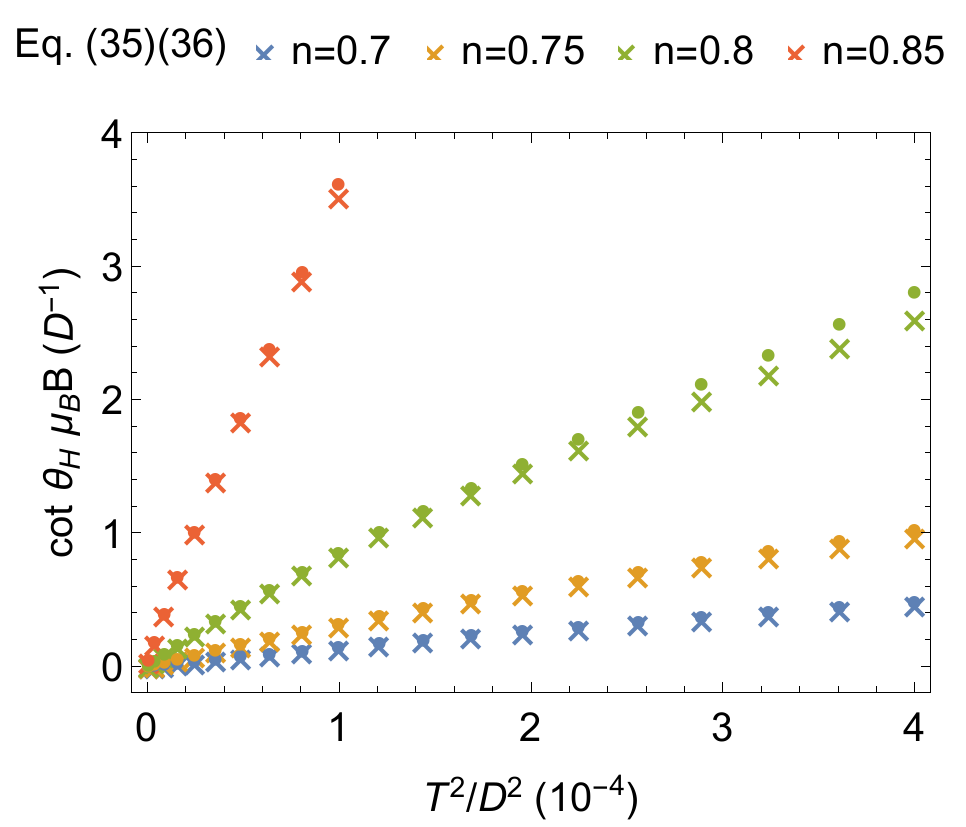}}
   \caption{$\alpha_{xx}$ (dashed lines) and $\alpha_{xy}$ (solid symbols) (\ref{subfig:tauh1-tauh}), $\rho_{xx}$ (\ref{subfig:rhoxx-star}), $\cot(\theta_H)$ (\ref{subfig:HA-star}) computed from Eq. (\ref{eq:tau-asymp}) and (\ref{eq:tau-h-asymp}) using ECFL results. The asymptotic values are denoted by crosses whereas the ECFL results of Eq. (\ref{eq:sxx-ep}) and (\ref{eq:sxy-ep}) are denoted by solid circles.}
   \label{fig:coeff-star}
 \end{figure*}

Now we keep $B(\omega,T)$ to $\mathcal{O}(\omega^2)$ and $A(\omega,T)$
to $\mathcal{O}(\omega)$, which are the lowest orders required to capture all important features of the exact results. Then Eq.~(\ref{eq:sigma-xx-asymp}) and (\ref{eq:sigma-xy-asymp}) can be simplified as
\begin{align}
  \begin{split}
 & \sigma_{xx} \simeq  \frac{\sigma_0 F^0_1}{D^2 B_0}   (D^2 - A_0^2)^{3/2} \Big(  1 - \frac{3 \pi^2F^2_2}{F_1^0}\frac{ T^2   A_0 A_1}{ \Delta (D^2 - A_0^2)}  \Big),
\end{split}
\label{eq:sxx-asymp}\\
  \begin{split}
    & \sigma_{xy}/B \simeq \frac{\sigma_0 q_e  F^0_2}{2  D^2 B_0^2}{ A_0 (D^2-A_0^2)^{3/2}}\\
    & \times \Big(1 +\frac{\pi^2 F^2_3}{F_2^0} \frac{ T^2   A_1 }{\Delta  A_0} (1-\frac{3 A_0^2}{D^2 - A_0^2}) \Big).
\end{split}
\label{eq:sxy-asymp}
\end{align}
The coefficients are defined as\cite{Fmn}
\begin{equation}\label{eq:F-mn}
F_m^n = \frac{\pi}{4} \int_{-\infty}^{\infty}  \frac{d x}{\cosh^2 (\pi x/2)} \frac{x^n}{(1+x^2)^{m}}.
\end{equation}
Using Eqs. (\ref{eq:sxx-asymp}), (\ref{eq:sxy-asymp}) and [\onlinecite{Fmn}], we can write
\begin{align}
  \sigma_{xx}& \simeq  \sigma_{xx, 0} (1 - \alpha_{xx}),\label{eq:tau-asymp}\\
  \sigma_{xy} &\simeq \sigma_{xy,0} (1 - \alpha_{xy}),\label{eq:tau-h-asymp}
\end{align}
with
\begin{align}
 & \sigma_{xx,0}  = \sigma_0  \frac{(D^2-A^2_0)^{3/2}}{D^2}\frac{0.822467 }{B_0}  , \\
 &\sigma_{xy,0}/B  =\sigma_0 q_e\frac{A_0 (D^2-A^2_0)^{3/2}}{D^2}\frac{0.355874}{B^2_0},\\
 & \alpha_{xx}  = \frac{A_1 A_0}{D^2 - A_0^2} \frac{3.98598 \times T^2}{\Delta}, \label{eq:alpha-xx}\\
 & \alpha_{xy}  = - A_1 \big(\frac{1}{A_0} - \frac{3A_0}{D^2 - A_0^2} \big)\frac{2.12075 \times T^2}{\Delta}\label{eq:alpha-xy}.
\end{align}
$\sigma_{xx,0}$ agrees with previous works\cite{Perepelitsky2016,Ding2017}. $\alpha_{xx (xy)}$ are relative corrections due to $\Delta$ and $A_1$ comparing to $\sigma_{xx(xy),0}$.
Numerical results of $\alpha_{xx}$ and $\alpha_{xy}$ are shown in Fig.~(\ref{subfig:tauh1-tauh}).
We find that $\abs{\alpha_{xx}}$ is less than $5\%$ even at the highest temperature. However, $\alpha_{xy}$ becomes $\mathcal{O}(1)$ in the GCSM regime.
Therefore, we obtain the following asymptotic $\tan \theta_H$ by omitting $\alpha_{xx}$:
\begin{align}
  \cot(\theta_H) \simeq \frac{\cot \theta_{H,0}}{ (1 - \alpha_{xy})}, \label{eq:theta-h-asymp2}\\
  \cot \theta_{H,0}/B = q_e \frac{B_0}{0.432691 A_0} .
\end{align}

We show $\rho_{xx}$ and $\cot(\theta_H)$ computed from the asymptotic expressions Eq. (\ref{eq:tau-asymp}) and (\ref{eq:tau-h-asymp}) in Fig. (\ref{fig:coeff-star}). The asymptotic values are denoted by crosses whereas the results of Eq. (\ref{eq:sxx-ep}) and (\ref{eq:sxy-ep}) are denoted by solid circles. The numerical results of Eq.~($\ref{eq:theta-h-asymp2}$) recover the second $T^2$ regime.

 {Therefore, we find that the $\alpha_{xy}$ term due to the higher order terms of $A(\omega, T)$ and $B(\omega,T)$ gives rise to the second $T^2$ regime of $\cot(\theta_H)$.
 Typically such correction is small, such as is the case of $\alpha_{xx}$.}
The significant difference between $\alpha_{xx}$ and $\alpha_{xy}$ is understood by examining Eq. (\ref{eq:alpha-xx}) and (\ref{eq:alpha-xy}) more closely.
% $F^0_{1(2)}(c)$ is only weakly dependent on $c$ even though $c$ goes up to $\sim 0.3$ at the highest temperature and can be approximated as $T$-independent constants. $F^1_{1}(c)$ and $F^1_{2}(c)$ are both $\propto c$.
Both $\alpha_{xx}$ and $\alpha_{xy}$ are $\propto A_1 T^2/\Delta$ with slightly different constant factors. Since $A_0 \ll D$ and almost independent of $T$, we can ignore the $3A_0(D^2-A_0^2)^{-1}$ term of $\alpha_{xy}$.
Hence the difference is mostly determined by a factor
\begin{equation}
\alpha_{xy} / \alpha_{xx} \sim A_0^{-2},
\end{equation}
which greatly enhances $\alpha_{xy}$.
% $\Phi^{xx\prime}[A(0,T)]/\Phi^{xx}[A_0] \propto A_0$ while $\Phi^{xy\prime}[A(0,T)]/\Phi^{xy}[A(0,T)] \propto A^{-1}(0,T)$ considering that $\abs{A(0,T)}$ is fairly small comparing to 1. Hence the $\alpha_{xy}$ is strongly amplified by the factor $A^{-1}(0,T)$.

{In the GCFL regime, $\alpha_{xy}$ is negligible, the coefficient of the $T^2$ behavior is
\begin{equation}\label{eq:QFL-asymp}
  Q_{FL} = \frac{B \cot(\theta_H)}{T^2} \simeq \frac{ \pi^3}{0.432691  \times q_e A_0 \Omega_{\Sigma}(T\to 0)}.
\end{equation}
$\Omega_{\Sigma}(T)$ is almost a constant in the GCFL regime hence approximated by its zero temperature value $\Omega_{\Sigma}(T \to 0)$\cite{fitting}.
In the GCSM regime, both $\Omega_{\Sigma}$ and $\Delta$ becomes linear-in-$T$:
\begin{align}
  & \Omega_{\Sigma}(T) \simeq \Omega_{0} (T + T_{\Omega}), \\
  & \Delta(T) \simeq \Delta_0 (T + T_{\Delta}),
\end{align}
where $\Omega_{0}(\Delta_0)$ and $T_{\Omega(\Delta)}$ are fitting parameters\cite{fitting}. %Only $T_\Omega$ varies from $T_\Omega > T_{FL}$ for $n=0.7$ to $ T_\Omega \ll T_{FL}$ for $n = 0.85$ while other fitting parameters remain larger than $T_{FL}$. Therefore,
By keeping only the constant term we obtain
\begin{equation}\label{eq:QSM-asymp}
 Q_{SM} \simeq  \frac{ \pi^3 }{0.432691  \times q_e A_0 \Omega_{0} (T_\Delta + T_\Omega)}.
\end{equation}
We compare the actual $Q_{FL}$ and $Q_{SM}$ with Eq.~(\ref{eq:QFL-asymp}) and (\ref{eq:QSM-asymp}) in Fig.~(\ref{fig:Qplot}).
\begin{figure}
  \centering
  \includegraphics[width=.85\columnwidth]{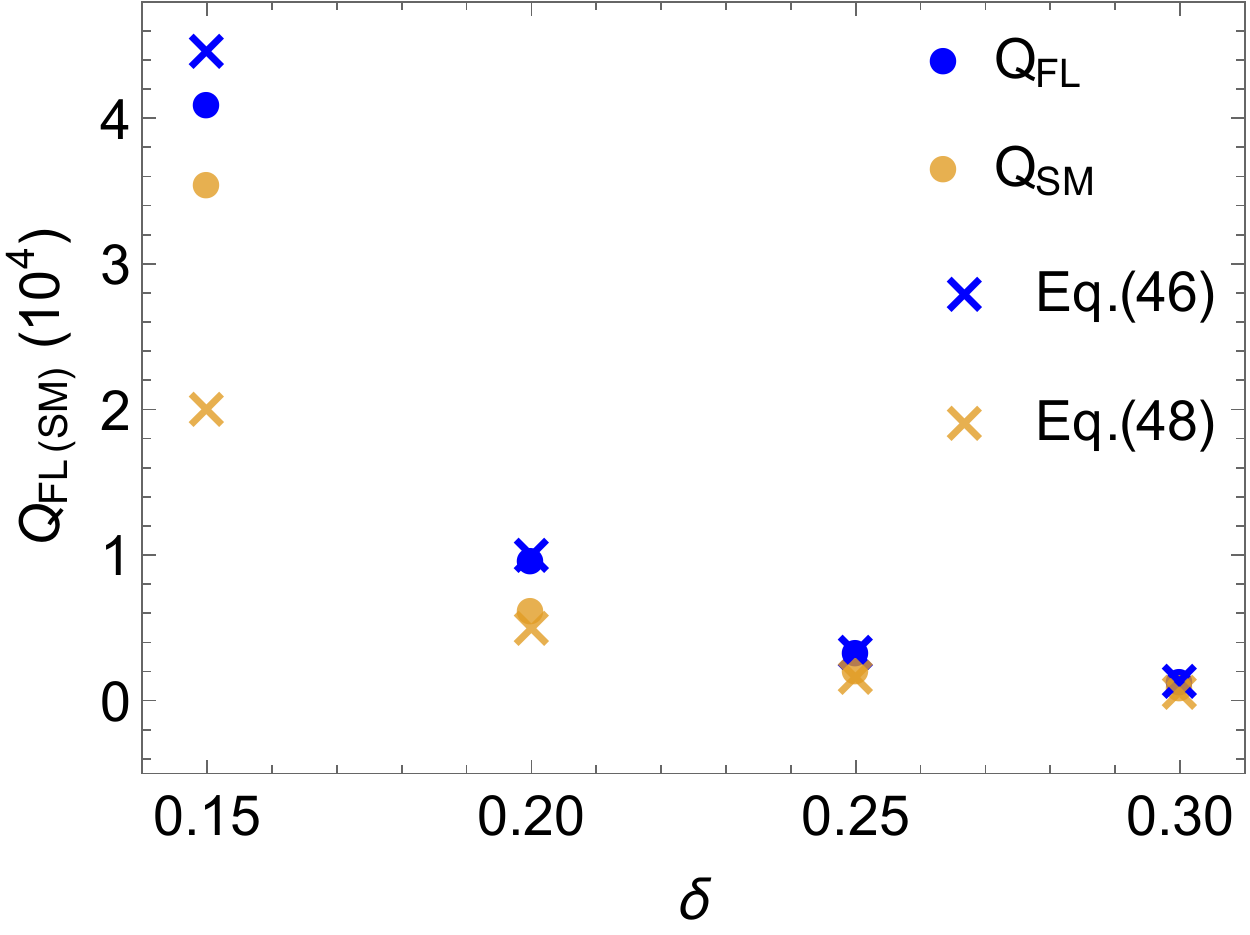}
     \caption{Eq.~(\ref{eq:QFL-asymp}) and (\ref{eq:QSM-asymp}) (crosses) compared with $Q_{FL}$ and $Q_{SM}$ (solid circles) obtained by fitting the exact $\cot(\theta_H)$.}
   \label{fig:Qplot}
 \end{figure}
}
%We see that, without $\alpha_{xy}$, $\cot \theta_{H,0} \propto B_0 \propto \rho_{xx}$ and would have the same $T$-dependence as $\rho_{xx}$ in both GCFL and GCSM regimes.
% \dwx{In the GCFL regime, $A_1 = Z^{-1}$ is independent of $T$ and $\Delta \propto T^{b}$ with $b<1$, therefore $\alpha_{xy}$}
%The $T$-dependence of $\alpha_{xy}$ is primarily determined by the $T$-dependence of $\Delta$ since both $A_0$ and $A_1$ are only weakly $T$-dependent.
%In the GCFL regime, $\alpha_{xy}$ is small and negligible.
%In the GCSM regime, $\alpha_{xy}$ becomes substantial and $\propto T^2/\Delta \propto T$ which alters $\cot(\theta_H)$ to be $\propto T^2$.

According to the above analysis, the second $T^2$ behavior of $\cot(\theta_H)$ is due to the combination of two things:
\begin{itemize}
\item the dynamic particle-hole anti-symmetric component of $\rho_{\Sigma}(\omega)$ characterized by the energy scale $\Delta$. Its contribution to transport becomes important when $\pi T$ becomes comparable to $\Delta$;
\item the particular form of the transverse transport function $\Phi^{xy}(\epsilon)$ that causes $\Phi^{xy\prime}[A_0]/\Phi^{xy}[A_0] \propto A^{-1}_0$. Without this factor, $\alpha_{xy}$ would be negligible as $\alpha_{xx}$. This particular form of $\Phi^{xy}(\epsilon)$ is due to the particle-hole symmetry of the bare band structure.
\end{itemize}

\section{Discussion}\label{sec:discussion}

We have shown that Hall constants, Hall angles, optical
conductivities, and optical Hall angles calculated by {ECFL agree
reasonably well with the DMFT results}. The differences tend to
increase at higher densities and higher temperatures as noted
earlier\cite{Perepelitsky2016}. 

We focused on the differences in the behavior above and below the
Fermi liquid temperature scale $T_{FL}$, i.e., from the GCFL regime to
the GCSM regime. Below $T_{FL}$, both $\rho_{xx}$ and $\cot(\theta_H)$
$\propto T^2$. Equivalently, $R_H$ has very weak $T$-dependence since
$R_H = \rho_{xx} / \cot(\theta_H)$. When $T>T_{FL}$, however,
$\cot(\theta_H)$ passes through a slight downward bend and continues
as $T^2$ whereas $\rho_{xx} \propto \, T$. The significance of the
downward bend is that it signals the crossover to the strange metal
regime from the Fermi liquid regime. 

We explored the long-standing two-scattering-rate problem by
calculating both the optical conductivities and optical Hall angles,
and the corresponding scattering rates. Below $T_{FL}$, both
$\sigma_{xx}(\omega)$ and $\tan \theta_H(\omega)$ exhibit Drude peaks,
which is a manifestation of transport dominated by quasiparticles. The
corresponding scattering rates can be extracted by fitting to the
Drude formula in the appropriate frequency range. Above $T_{FL}$, the
Drude peak for $\sigma_{xx}(\omega)$ becomes broadened, i.e.,
$\sigma_{xx}(0)/\sigma_{xx}(\omega) - 1 \sim \omega^2$ for an even
larger range that keeps growing with increasing temperature. In this
case, fitting to the Drude formula is still valid, and the scattering
rate shows similar trends as a function of temperature as the $dc$
resistivity. For $\theta_H(\omega)$, the Drude peak range {is very
narrow}, but nonetheless persists for all temperatures that we study
in this work. Similarly, the extracted scattering rate $\Gamma_H$
shows similar trends as a function of temperature as the $dc$ Hall
angle. At lower dopings and higher temperatures, it seems possible
that the Drude peaks of $\theta_H(\omega)$ would disappear and the
fractional power law would stretch down to nearly $\omega = 0$.

By comparing the two optical scattering rates through their ratio,
$\Gamma_H/ \Gamma_{tr}$, we clearly demonstrated that $\Gamma_H$ and
$\Gamma_{tr}$ are equivalent below $T_{FL}$, but that they quickly
become two distinguishable quantities when the system crosses over
into the strange-metal region.

By carefully examining the asymptotic expressions of $\sigma_{xx}$ and
$\sigma_{xy}$ we established that the different temperature dependence
of $\cot(\theta_H)$ in the GCSM regime is governed by a correction
caused by both the dynamical particle-hole anti-symmetric component of
$\rho_\Sigma(\omega)$ and the particle-hole symmetry of the bare band
structure. This correction is turned on when $T$ becomes comparable to
$\Delta$, the characteristic energy scale of the anti-symmetric
components of $\rho_\Sigma(\omega)$.

It would be useful to examine the bend in $\cot(\theta_H)$ more
closely in experiments in cuprate materials, where such a feature is
apparently widely prevalent but seems to have escaped comment so far.
In particular, one would like to understand better if the longitudinal
resistivity and the cotangent Hall angle show simultaneous signatures
of a crossover, as the theory predicts.

\section{Acknowledgements}
The work at UCSC was supported by the U.S. Department of Energy (DOE), Office of Science, Basic Energy Sciences (BES) under Award \# DE-FG02-06ER46319.
R\v{Z} acknowledges the
financial support from the Slovenian Research Agency (research core
funding No.~P1-0044 and project No.~J1-7259).

%\bibliography{./library}

\begin{thebibliography}{23}
\bibitem{Fmn} Here we give numerical values of $F_{m}^{n}$ here,
  $F^0_1=\pi^2/12 =0.822467,\ F^2_2=\pi^2/24 -  \zeta(3)/4= 0.110719,
 \ F_2^0=\pi^2/24 +  \zeta(3)/4 = 0.711748,  \ F_3^2=\pi^2/96 -\pi^4/960+  \zeta(3)/16  =0.0764691,
 $ where $\zeta(3)=1.20206\ldots$ is the Reimann zeta function.
\bibitem{fitting}
  Here we give numerical values of $\Omega_{\Sigma}(T\to 0)$, $\Omega_0,\ \Delta,\ T_{\Omega}$ and $T_\Delta$ for $n=0.7,\ 0.75,\ 0.8,\ 0.85$ respectively:
  \begin{tabular}{lllll}
    \hline
 n &0.7  & 0.75 & 0.8 & 0.85 \\
 $\Omega_{\Sigma}(T\to 0)$ & 0.194326 & 0.10234 & 0.0443113 & 0.0135004\\
 $\Omega_0$ & 5.99932 & 4.97066 & 3.79492 & 2.52377    \\
  $\Delta$ & 5.15313 & 5.58921  & 5.97819 &  6.05317\\
  $T_\Omega$ &0.0257944 & 0.0143857 & 0.00607066 &0.000456418 \\
   $T_\Delta$ & 0.0346638 & 0.0248982 & 0.0171223 & 0.0118911
\end{tabular}
\end{thebibliography}

\end{document}